\documentclass[%
 aip,
 jcp,%
 amsmath,amssymb,
 reprint,%
 onecolumn,
showkeys,
floatfix
]{revtex4-1}

\usepackage{graphicx}
\usepackage{dcolumn}
\usepackage{bm}
\usepackage{gensymb}
\usepackage{placeins}
\usepackage{natbib}
\usepackage{multirow}
\usepackage{float}
\usepackage{siunitx}
\usepackage{units}
\DeclareSIUnit\calorie{cal}
\DeclareSIUnit\kcal{\kilo\calorie}
\usepackage[version=3]{mhchem}
\begin{document}


\title[Embedding parameters in \textit{ab initio} theory]{Embedding parameters in \textit{ab initio} theory to develop approximations based on molecular similarity}

\author{Matteus Tanha}
\author{Haichen Li}
\affiliation{
Department of Chemistry, Carnegie Mellon University, Pittsburgh, PA 15213
}
\author{Shiva Kaul}
\affiliation{
Machine Learning Department, Carnegie Mellon University, Pittsburgh, PA 15213
}
\author{Alexander Cappiello}
\affiliation{
Department of Chemistry, Carnegie Mellon University, Pittsburgh, PA 15213
}
\author{Geoffrey J. Gordon}
\affiliation{
Machine Learning Department, Carnegie Mellon University, Pittsburgh, PA 15213
}
\author{David J. Yaron}
\email{Yaron@cmu.edu}
\affiliation{
Department of Chemistry, Carnegie Mellon University, Pittsburgh, PA 15213
}

\thanks{Funded by the National Science Foundation through grant numbers 1027985 and 1135553.}

\date{\today}

\begin{abstract}

A means to take advantage of molecular similarity to lower the computational cost of electronic structure theory is explored, in which parameters are embedded into a low-cost, low-level (LL) \textit{ab initio} model and adjusted to obtain agreement with results from a higher-level (HL) \textit{ab initio} model. A parametrized LL (pLL) model is created by multiplying selected matrix elements of the Hamiltonian operators by scaling factors that depend on element types. Various schemes for applying the scaling factors are compared, along with the impact of making the scaling factors linear functions of variables related to bond lengths, atomic charges, and bond orders. The models are trained on ethane and ethylene, substituted with -\ce{NH2}, -\ce{OH} and -\ce{F}, and tested on substituted propane, propylene and t-butane. Training and test datasets are created by distorting the molecular geometries and applying uniform electric fields. The fitted properties include changes in total energy arising from geometric distortions or applied fields, and frontier orbital energies. The impacts of including additional training data, such as decomposition of the energy by operator or interaction of the electron density with external charges, are also explored. The best-performing model forms reduce the root mean square (RMS) difference between the HL and LL energy predictions by over 85\% on the training data and over 75\% on the test data. The argument is made that this approach has the potential to provide a flexible and systematically-improvable means to take advantage of molecular similarity in quantum chemistry.

\end{abstract}

\pacs{31. Electronic structure of atoms and molecules: theory}

\maketitle

\section{Introduction}\label{sec:intro}

\noindent \textit{Ab initio} electronic structure theory has made great strides over the past few decades in developing methods with systematically improvable accuracy. For many classes of molecules, chemically accurate predictions can now be obtained\cite{curtiss2007,gordon2009}. A remaining challenge is lowering the computational cost such that accurate predictions may be obtained on large systems, such as those of interest in biological and materials applications. Molecular systems have two properties that provide the basis for approximations that lower computational costs: nearsightedness and molecular similarity. Nearsightedness implies that interactions become simpler at long-range\cite{prodan2005}, and can be replaced by increasingly coarse-grained multi-polar interactions. Algorithms that take full advantage of nearsightedness can achieve a computational cost that scales linearly with system size in the limit of large systems\cite{Goedecker1999,VanderVaart2000,Yang1991,Greengard1987,White1996,Shang2010}.  Another aspect of molecular systems that can lead to a substantial reduction in computational cost is molecular similarity. Similarity implies that molecular fragments such as functional groups behave similarly in similar environments. Molecular similarity is the basis for molecular mechanics\cite{brooks2009} and semiempirical quantum chemistry (SEQC)\cite{ridley1973,dewar1990,repasky2002,thiel2014}. While such methods have had great success, the approximation schemes are not as well controlled as those of nearsightedness. The work reported here explores approaches aimed at increasing the level of control in approximations based on molecular similarity.

Approximations based on nearsightedness benefit from having a clear physical basis that inspires the form of the approximation. For example, when computing Coulomb interactions at long range, charge distributions may be replaced with multipoles. This physical basis suggests specific forms for the approximations of linear scaling\cite{Goedecker1999} and fragment-based methods\cite{gordon2009}. The resulting algorithms may have parameters that define, for example, the degree to which interactions are simplified at long range. These parameters can be adjusted to find an optimal balance between computational savings and errors introduced by the approximation. This high level of control is related to the direct physical basis underlying the approximation. 

Achieving a similar level of control in approximations based on molecular similarity is more challenging. Ideally, the resulting methods would provide a means to smoothly improve the accuracy of the approximation such that an appropriate balance between accuracy and computational cost could be found for the target application. 

The flexible model forms of machine learning, such as neural nets, enable construction of models that are trained to data on representative molecules and then used to make predictions for molecules not included in the training set. Such models have been successfully used to develop corrections to predictions of \textit{ab initio} theory that are applicable across broad classes of molecules and that substantially reduce the errors in heats of formation and other properties~\cite{Hu2003,Duan2004,Li2007,Sun2014,Gao2009,Gao2009a,Wu2008,Wu2010,Balabin2009,Ediz2009}. Models have also been developed that generate predictions using descriptors, such as the Coulomb matrix\cite{Rupp2012}, that do not require solution of the Schr\"{o}dinger equation. The resulting models produce results that compete with quantum chemistry at substantially reduced computational cost\cite{Rupp2012,Hansen2013,Montavon2013}. 

The current work explores the use of model Hamiltonians as flexible functional forms for model development, and so bridges between the flexible model forms of machine learning\cite{Mitchell1997}, such as neural nets or reproducing-kernel Hilbert space methods, and the model Hamiltonians of SEQC, such as Intermediate Neglect of Differential Overlap (INDO)\cite{ridley1973} and Neglect of Diatomic Differential Overlap (NDDO)\cite{dewar1990}.  As in machine learning, the goal is to create model forms that are sufficiently flexible that patterns in the training data may be discovered. As in SEQC, the strategy is to use \textit{ab initio} quantum chemistry as inspiration for model forms that can well describe chemical phenomena.

The approach used in the investigations is adapted from machine learning. First, a set of molecular systems across which similarity may be reasonably expected to hold is specified. This is done by selecting a group of related molecules and creating a set of molecular ``instances'' by distorting the molecular geometry and applying external electrostatic environments. Predictions of a high-level (HL) \textit{ab initio} model are then generated to serve as training data. The parameters of a lower-cost model are adjusted to reproduce the training data and the performance of the model is evaluated on instances not included in the training data.

The approach used to create the model is adapted from traditional SEQC: we embed empirical parameters into a quantum chemical Hamiltonian. First, matrix elements of the operators that appear in a low-level (LL) \textit{ab initio} Hamiltonian are generated, including kinetic energy, attraction between the electrons and each nucleus, and the electron-electron repulsion. Selected matrix elements of these operators are then multiplied by scaling factors that serve as the parameters of the model, leading to a parametrized low-level (pLL) model. Matrix elements that are not scaled retain the values of the LL \textit{ab initio} Hamiltonian. This general approach is flexible because a variety of LL Hamiltonians may be envisioned into which parameters may be embedded. For these initial studies, the LL model is restricted to self-consistent-field (SCF) solutions within a minimal basis set. A variety of schemes for embedding parameters can also be envisioned. Here, four different approaches to applying the scaling factors are explored along with the impacts of making the scaling factors sensitive to the molecular context. 

A number of factors point to the ability of a parametrized minimal-basis Hamiltonian to provide an accurate description of chemical systems. For one, such Hamiltonians can qualitatively describe a broad range of chemical phenomena, including valency, bond formation and aromaticity. Existing SEQC methods have had considerable success in using such model forms to achieve quantitative accuracy\cite{ridley1973,dewar1990}. In addition, at the SCF level, a minimal basis Hamiltonian may be found that, for a specific molecular instance, generates the electron density and related molecular properties of a large-basis Hamiltonian\cite{quambo}. In the studies reported here, both the HL and pLL model use SCF solutions such that  the models differ only in the use of a larger (6-31G**) basis set for the HL model.  A minimal-basis Hamiltonian should therefore provide a model form with sufficient flexibility to reproduce the HL predictions.

Model Hamiltonians also have the potential advantage of enabling models trained on small molecules to be applicable to larger molecules. By training to the energy of small molecules, the model incorporates information on short-range interactions. By training to the charge distribution of small molecules, the model may incorporate information on longer-range interactions. The pLL models used here support this by using the LL form for those portions of the Hamiltonian that describe long-range interactions.  At sufficiently long range in SCF calculations, the interaction between regions of electron density becomes equal to the Coulomb interaction between those regions. In the pLL models, the electron density, $\rho \left( {\bf{r}} \right)$, is described by the density matrix expressed in the LL basis set,
\begin{equation}\label{eq:rhobasis}
\rho \left( {\bf{r}} \right) = \sum\limits_{i,j} {{\rho _{i,j}}{\phi _i}\left( {\bf{r}} \right){\phi _j}\left( {\bf{r}} \right)},
\end{equation}
where $\rho _{i,j}$ is the density matrix and $\phi_i\left( {\bf{r}} \right)$ are LL basis functions. Because integrals describing longer-range interactions retain their LL \textit{ab initio} values, the longer-range interactions in the pLL models have the form
\begin{equation}\label{eq:rhorhobasis}
{E_{long\,range}} = \sum\limits_{\scriptstyle i,j \in regio{n_1}\hfill\atop
	\scriptstyle k,l \in region{_2}\hfill}^{} {{\rho _{i,j}}\left( {ij|kl} \right){\rho _{k,l}}}
\end{equation}
where $\left( {ij|kl} \right)$ are two-electron integrals in chemist's notation. Eq.~\ref{eq:rhorhobasis} is an accurate expression for the interaction between the charge distributions of Eq.~\ref{eq:rhobasis}. The degree to which the pLL model provides an accurate description of longer-range interactions is then related to the degree to which the density of Eq.~\ref{eq:rhobasis} accurately describes the HL density. This accuracy is limited by both the use of a LL basis and by the degree of success obtained in training the model parameters to best reproduce the HL charge density. The studies below test the ability of models trained on substituted ethane and ethylene to make predictions for larger systems. 

Despite some similarities with traditional SEQC, substantial differences arise from the manner in which the parameters are embedded in the LL \textit{ab initio} model. Here, when all embedded parameters are zero, the scaling factors become one and the pLL model returns to the LL model. This differs substantially from traditional SEQC where the models do not map as directly onto an underlying LL \textit{ab initio} method. For example, in INDO and NDDO methods\cite{ridley1973,dewar1990}, the one-electron tight-binding terms between atoms, $\beta$, are taken as proportional to overlap integrals between atomic orbitals, as opposed to being based on integrals of the operators that contribute to these bonding interactions. 

SEQC models also make strong approximations regarding the interactions between charge densities on different atoms by including only one-center and two-center two-electron integrals. In CNDO and INDO, the form used for the two-center integrals essentially reduces the Coulomb interactions between atoms to that between atomic charges. In NDDO, the form of the interactions includes higher order multipoles but is still based on a model form that does not map directly onto a LL \textit{ab initio} method. From the perspective adopted here, such model forms combine approximations related to nearsightedness with approximations related to molecular similarity, by using empirical parameters to compensate for the use of simplified multipolar interactions at closer range than is likely justifiable based solely on nearsightedness. The pLL model forms explored here take a different approach by attempting to smoothly transition from local interactions, which are empirically modified through scaling factors, to more remote interactions, which are unscaled and thus retain the form of the LL \textit{ab initio} model, Eq.~\ref{eq:rhorhobasis}. 

SEQC models also do not include core electrons and instead use empirical core-core interaction terms, which are helpful in improving model predictions. In density functional tight binding (DFTB), such core-core interactions are the only terms treated empirically, with the electronic Hamiltonian being computed from DFT\cite{Elstner1998,Cui2001,Elstner2006,Gaus2012}. Here, the pLL model includes core electrons through basis functions that are identical to those used for the core electrons in the HL model against which the pLL model is trained. The pLL model does not scale matrix elements involving core electrons, nor are empirical core-core interaction terms added to the model. This is done to prevent effects that arise from valence electrons from being absorbed into core-core interaction terms.

A number of means are explored to improve training of the model. That the parameters of the pLL model are associated with molecular fragments has the advantage of allowing model parameters to be transferred between different molecules. However, challenges in training the model result from the parameters being associated with molecular fragments while model performance is measured on the molecule as a whole. The training data includes total energy and frontier orbital energies as target properties the model is intended to predict. In addition, the extent to which inclusion of additional properties can enrich the training data and so lead to better model performance is explored. These additional properties include decomposition of the total energy by operator into kinetic energy, electron-nuclear and electron-electron components. Inclusion of data regarding the molecular charge distribution is also explored, to include information related to longer-range interactions, Eq.~\ref{eq:rhorhobasis}.

Section~\ref{sec:data} describes the data used to train and test the pLL models. Section~\ref{sec:params} describes the various approaches used to apply scaling factors to the operator matrix elements of the LL Hamiltonian. The manner in which these scaling factors are made functions of the molecular context is described in Section~\ref{sec:context}. Training of pLL models is described in Section~\ref{sec:fitting} and the results are discussed in Section~\ref{sec:results}. Section~\ref{sec:discussion} reflects on the outcomes and discusses possible next steps. 

\section{Chemical data}\label{sec:data}
\noindent The data used to train and test the model is constructed by selecting a set of molecules across which molecular similarity may be expected to hold, generating geometries of these molecules that are distorted from the minimum-energy geometry in some controlled manner, and finally placing these molecules in external electrostatic environments. The combination of a molecular geometry and an environment will be referred to as a molecular ``instance''.

The geometries are distorted in a manner that is meant to provide uniform sampling over some prespecified range of perturbations to both bond lengths and bond angles. The approach utilizes a function, ${\bf{R}}\left( {\bf{\xi }} \right)$, that maps the internal coordinates of the z-matrix, $\xi$, to all bond lengths and bond angles, $\bf{R}$, of the molecule. A set of target bond lengths and angles is generated as 
\begin{equation}\label{eq:Rtarget}
{{\bf{R}}^{{\rm{target}}}} = {\bf{R}}\left( {{{\bf{\xi }}^{opt}}} \right) + {\bf{X}}
\end{equation}
where $\bf{\xi}^{opt}$ are the internal coordinates of the optimized structure and $\bf{X}$ is a list of random numbers that specify distortions in bond lengths and angles. A uniform random distribution is used for $\bf{X}$ to ensure that the geometries well sample the specified range of bond lengths and angles. Because there are typically more bond lengths and angles than internal coordinates, ${\bf{R}}\left( {\bf{\xi }} \right)$ is not directly invertible. The internal coordinates are therefore adjusted to give the best match to ${{\bf{R}}^{{\rm{target}}}}$ by minimizing the following root mean square (RMS) deviation,
\begin{equation}\label{eq:geomMin}
\sum\limits_{i = 1}^N {{{\left[ {{w_i}\left( {{R_i}\left( {\bf{\xi }} \right) - R_i^{{\rm{target}}}} \right)} \right]}^2}},
\end{equation}
where $N$ is the total number of bond lengths and angles and $w_i$ is a weight that is inversely proportional to the width of the uniform random number distribution used in generating ${{\bf{R}}^{{\rm{target}}}}$.

In cases where a dihedral angle between groups is sampled, this is done by creating a z-matrix where a single internal coordinate controls the relative rotation of the two groups. This coordinate is assigned to a randomly generated value and not altered during the minimization of Eq.~\ref{eq:geomMin}. For propane, propylene and butane, the range of dihedral angles is restricted to prevent overlapping groups.  

To include information on the manner in which the systems respond to external electrostatic environments, each molecule is placed in static electric fields applied along the X, Y and Z directions. A magnitude of 0.00333 a.u. was chosen, which is sufficient to induce a dipole of 0.1 Debye in water. Each molecular geometry thus leads to four instances, one instance in zero field and three instances for fields along the X, Y and Z directions.  

Section~\ref{sec:intro} argued that the degree to which long-range interactions are well described by the pLL model is related to the degree to which the density of Eq.~\ref{eq:rhobasis} agrees with that of the HL model. To obtain a measure of electron density that is comparable across basis sets, the interaction of the electron density with a set of point charges surrounding the molecule, $E_{chg}$, is computed. One hundred locations for the point charges are first constructed using the Chelpg method\cite{Breneman1990}. Values are then assigned to these charges from a uniform random distribution of -0.08~$e$ to +0.08~$e$. Twenty such external point charge environments are created for each molecular geometry.  

For each instance in the dataset, the following properties are generated.
\begin{description}
\item[$\bf{E_{tot}}$] The total energy of the molecule, without inclusion of the interaction with an external field. The external field thereby serves to perturb the electron density and $E_{tot}$ measures the electronic energy associated with this perturbation. 
\item[$\bf{E_{orb}}$] The energy of the HOMO and LUMO orbitals. Because the calculations are being done at the SCF level, these are Koopman's theory estimates of the ionization potential and electron affinity.
\item[$\bf{KE}$, $\bf{EN_A}$, $\bf{E_2}$] Expectation values of the operators that make up the Hamiltonian. $KE$ is the total kinetic energy. $EN_A$ is the interaction of the electron density with the $A^{th}$ nucleus, and $E_2$ is the total electron-electron repulsion energy. Unlike decomposition of the energy into atoms or molecular fragments, the decomposition of the energy by operator is uniquely defined and directly comparable across basis set.
\item[$\bf{E_{chg}}$] The interaction energy between the electron density and the randomized point charges described above.  
\end{description}
The model parameters are adjusted to obtain agreement between the pLL and HL models for the above quantities, as described in Section~\ref{sec:fitting}.

The following data sets were generated. Models were trained on the first three datasets, with the remaining used for model testing. 
\begin{description}
\item[\textit{ethane}] Includes all 16 unique ways to place between zero and two substituents, selected from -\ce{NH2}, -\ce{OH}, and -\ce{F}, on ethane. The optimized geometry, at the 6-31G level,  is included along with distorted geometries created as described above. The bond lengths are uniformly distorted by $\pm$0.2 \AA, bond angles are distorted by $\pm$10\degree, and rotation about the central C-C bond is randomized. For each such geometry, four instances are created corresponding to no external field, and external fields along the X, Y and Z axes of 0.0033 atomic units. A training and validation set are created, both of which contain the optimized geometry along with 10 distorted geometries in the training set and 9 in the validation set, leading to 640 instances in the training and 576 instances in the validation set.
\item[\textit{ethylene}] This dataset is generated in a manner identical to the \textit{ethane} dataset, except that the substituents are attached to ethylene and the bond lengths are uniformly distorted by $\pm$0.15 \AA. Just as in ethane, the geometries include full sampling of rotation about the central carbon-carbon double bond. The bond length distortions were reduced to make the disagreements between the initial LL and the HL model for $E_{tot}$ comparable. 
\item[\textit{combined}] A combination of the \textit{ethane} and \textit{ethylene} datasets. 
\item[\textit{propane}] Includes all unique combinations of zero to two of the above substituents on propane, but with at most one substituent attached to each carbon. The bond lengths are distorted by $\pm$0.2 \AA, bond angles are distorted by $\pm$10\degree, and rotation about one carbon-carbon bond is randomized subject to avoiding overlapping groups. A total of 576 instances are included in this dataset. 
\item[\textit{propylene}] This dataset is generated in a manner similar to the propane dataset, except that the substituents are attached to propylene, the bond lengths are uniformly distorted by $\pm$0.15 \AA, and rotation about the double bond is randomized. The substituents are either on each end of the molecule or adjacent across the double bond.
\item[\textit{t-butane}] This dataset is identical to the \textit{ethane} dataset but has two methyl groups attached to one of the carbons. This tests the ability of a model trained on molecules in which carbon has at most three heavy atom substituents to transfer to molecules in which carbon has four heavy atom substituents. This dataset consists of 576 instances.
\end{description}
The magnitude of geometric distortions used above was chosen to provide data that is challenging to model yet is on molecules of sufficiently small size that models can be trained in an efficient manner. The average change in energy associated with the distortions in the \textit{combined} dataset is $\SI{65}{\kcal\per\mole}$. Table~\ref{tab:SEcombined} shows disagreements between the HL model and traditional SEQC models for the \textit{combined} dataset. The relatively large values point towards the level of difficulty associated with these datasets, although this is only qualitative because the differences are quoted with respect to the HL model while the SEQC models are parametrized to experimental data.  
\begin{table}[H]
\centering
\begin{tabular}{cccc}
\hline
Method & E$_{tot}$ & E$_{HOMO}$ & E$_{LUMO}$\\
& $\SI{}{\kcal\per\mole}$ & eV & eV \\ 
\hline
CNDO/2 & 58.97 & 2.98 & 1.62\\
INDO   & 59.94 & 2.43 & 1.56\\
MNDO   & 34.49 & 1.99 & 3.55\\
AM1    & 32.95 & 1.84 & 3.37\\
PM3    & 34.81 & 2.09 & 3.78\\
\hline
\end{tabular}
\caption{RMS differences between the HL model and SEQC methods for the \textit{combined} dataset. E$_{tot}$ refers to the change in energy associated with geometric distortions and applied fields (see Section~\ref{sec:fitting}).}
\label{tab:SEcombined}
\end{table}

The amount of training data for the \textit{ethane} and \textit{combined} datasets is shown in Table~\ref{tab:outputs}. The values that most influence the training are E$_{tot}$ and E$_{orb}$. With the default weighting of E$_{chg}$ in the objective function of Section~\ref{sec:fitting}, E$_{chg}$ does not play a strong role in the fitting. The effects of increasing the weighting of E$_{chg}$ are explored in Section~\ref{sec:results}. Comparisons are also made between models trained with and without inclusion of the decomposed energies (KE, EN$_A$ and E$_2$). 

\begin{table}[h]
\centering
\begin{tabular}{ccc}
\hline
 & \textit{ethane} & \textit{combined} \\
 \hline
 E$_{tot}$  & 688 & 1376 \\
 E$_{orb}$ & 1408 & 2816 \\
 E$_{chg}$ & 13376 & 26752 \\
 KE & 688 & 1376\\
 EN$_A$ & 6665 & 11954 \\
 E$_2$ & 688 & 1376\\
 \hline
\end{tabular}
\caption{Amount of training data in the datasets used to develop pLL models.}
\label{tab:outputs}
\end{table}

\section{Embedding parameters in the low-level Hamiltonian}\label{sec:params}
\noindent Flexible model forms are created by embedding parameters into a low-level (LL) \textit{ab initio} model, to form a parametrized low-level (pLL) model. The parameters of the pLL model are then adjusted to obtain agreement with a higher-level (HL) \textit{ab initio} model. We have chosen to use SCF solutions for both the pLL and HL model, such that the empirical parameters are used only to compensate for differences related to the basis set.  For the HL model, a split-valence basis set with polarization functions (6-31G**) is used. For the valence electrons of the pLL model, the  STO-3G minimal basis is used. For the core electrons of the pLL model,  the core orbitals of the 6-31G basis are used, in order to make the description of the core electrons in the pLL model equivalent to that in the HL model. This combination of 6-31G core functions and STO-3G valence functions will be referred to as the modified STO-3G basis (mSTO-3G). The unparametrized low-level (LL) model therefore corresponds to SCF solutions within the mSTO-3G basis.

To initialize the pLL model, the matrix elements of each of the operators appearing in the electronic Hamiltonian are evaluated in the mSTO-3G basis of the LL model
\begin{equation}\label{eq:stoElements}
KE_{i,j} \ \  EN^A_{i,j}\ \  \left(ij|kl\right) \ \  H^{env}_{i,j} 
\end{equation}
where $KE$ is the kinetic energy operator, $EN^A$ is the interaction of the electrons with the $A^{th}$ nucleus, $\left(ij|kl\right)$ are the two-electron integrals in chemists notation, and $H^{env}$ is the interaction with the external electrostatic environment. The pLL model embeds parameters into the mSTO-3G Hamiltonian by multiplying subblocks of these matrices by scaling factors, $S$. Different scaling factors are used for different operator types ($KE$, $EN_A$, $E_2$) and for differing elements (H, C, N, O, F) involved in the subblock being modified. The scaling factors depend on context through the form
\begin{equation}\label{eq:SFactor}
  S = 1 + p_0 + \sum_{i=1}^{N_c} p_i c_i
\end{equation}
where, $c_i$, are context variables that describe the local environment of the atom or bond associated with the subblock being modified (Section~\ref{sec:context}). The total number of parameters, $p_i$, is then related to both the number of scaling factors and the degree to which these are made context sensitive, $N_c$. A number of different approaches, or policies, are explored for embedding the scaling factors. Note that matrix elements that are not scaled retain their LL values, and scaling is applied only to matrix elements between valence orbitals.

For diagonal blocks of one-electron operators, $O_{i,j}$ where $i$ and $j$ are orbitals on the same atom, one scaling factor is used for the $s$ orbitals and a different scaling factor is used for the subblock corresponding to the $p$ orbitals. For off-diagonal blocks of one-electron operators, $O_{i,j}$ where $i$ and $j$ are orbitals on different atoms, scaling is applied only when $i$ and $j$ are on bonded atoms. Two different policies are used for these off-diagonal blocks. In the default policy, different scaling factors are used for matrix elements connecting $s$ with $s$, $p$ with $p$, and $s$ with $p$ orbitals, leading to three parameters for each pair of elements. The $\sigma\pi$ policy adds an additional scaling factor per pair of heavy elements by using different parameters for $\sigma$ versus $\pi$ orientations of the $p$ orbitals. (This is done by temporarily rotating the $p$ orbitals such that one $p$ orbital of each atom lies along the internuclear axis.) For the attraction between the electron density and the $A^{th}$ nucleus, $EN^A_{i,j}$ is modified according to the above rules only when $i$, $j$, or both $i$ and $j$ reside on atom $A$.  $H^{env}$ is not modified. 

For electron-electron interactions, two different policies are explored. The \textit{2elec} policy modifies the two-electron integrals, $(ij|kl)$, directly. Scaling factors are applied only to those integrals that are retained in the NDDO approximation\cite{dewar1990} but, unlike NDDO, the remaining integrals retain their LL values instead of being set to zero.  For on-atom integrals, where $i$, $j$, $k$, and $l$ are on the same atom, the three parameter form developed by Slater~\cite{ridley1973,pople1967,slater1960} is used, which leads to parameters $F_0$, $G_1$, and $F_2$ per heavy element. The unscaled values of $F_0$, $G_1$ and $F_2$ are first determined from the two-electron integrals of the LL basis, and these values are then multiplied by scaling factors, leading to one scaling factor for hydrogen and three scaling factors for each heavy element. Between bonded atoms, all integrals $(ij|kl)$ where $i$ and $j$ are on one atom while $k$ and $l$ are on the other atom, are multiplied by a single scaling factor, leading to one scaling factor for each pair of elements. A scaling factor between non-bonded atoms is included only for non-bonded hydrogen atoms closer than 2.8 \AA, because this leads to a small improvement in model predictions. This nonbonded scaling uses bond order as its single context variable, $c_1$ of Eq.~\ref{eq:SFactor}. 

The \textit{JK} policy is an alternative to the \textit{2elec} policy based on scaling of the Coulomb, J, and exchange, K, matrices of the Fock operator. The scaling is applied in each iteration of the self-consistent solution of the Roothan equations. The unscaled two-electron integrals of the LL model are first used to evaluate the J and K matrices. The policies discussed above for the one-electron operators are then used to scale the J and K matrices, with different scaling factors used for J versus K.

\section{Context Sensitive Scaling Factors}\label{sec:context}
\noindent A given set of scaling factors is likely to be valid only over some limited range of molecules. To extend this range, the scaling factors are made functions of the current context of the atom or bond by making them linear functions of context variables, $c_i$ of Eq.\ref{eq:SFactor}. In training of the model parameters (Section~\ref{sec:fitting}), the context sensitivity is turned on sequentially in the order shown in Table~\ref{tab:contexts}. The first level of context involves only bond lengths and so only introduces simple geometry dependence into the scaling factors. The second and third level of context go beyond geometry to include aspects of the electronic structure of the molecule through atomic charges and bond orders. As charge is pushed onto (or pulled from) an atom, we expect the charge density to expand (or contract). This effect is not present in the minimal basis of the LL model. Making the scaling factors functions of atomic charge and bond orders is an attempt to compensate for this. For scaling factors applied to the diagonal, on-atom, matrix elements, the atomic charge seems most relevant and so this is included at the second level of context sensitivity. For scaling factors applied to off-diagonal blocks, the bond order seems more relevant and so this is included at the second level of context. The third level crosses these, using bond order for diagonal blocks and atomic charges for off-diagonal blocks. The context variables are calculated using the LL model.

\begin{table*}[!htbp]  

\begin{tabular}{@{}llll@{}}
\hline
& level & on-atom & between-atom\\ 
\hline
 &c1\footnote{$r$ refers to deviations of the bond length from 1.1\AA\ for C-H bonds and 1.5\AA\ for all other bonds.} & r: Average bond length to bonded atoms & r: Bond length\\
 &c2\footnote{$q$ refers to deviations from average values of 0.112 for H, -0.335 for C, -0.713 for N, -0.639 for O, and -0.417 for F.} 
 & q: Mulliken charge on the atom & bo: Bond order \\
 &c3 & bo: Average bond order to bonded atoms & q: charge difference between atoms\\ 
 \hline
\end{tabular}
\caption{Context variables, $c_i$ of Eq.\ref{eq:SFactor}, for the three levels of context added sequentially to the model during training.}
\label{tab:contexts}
\end{table*}

\section{Training the pLL Model}\label{sec:fitting}
\noindent The parameters of the pLL model are trained by minimizing
\begin{equation}\label{eq:obj}
Obj\left(\bf{p}\right) = \sum_i^{N_{data}} w_i^2 \left( X_i^{HL} - X_i^{pLL}\left(\bf{p}\right) \right)^2 + R\left(\bf{p}\right)
\end{equation}
where $Obj$ is the objective to be minimized, $\bf{p}$ is a vector containing all parameters in the scaling factors of Eq.~\ref{eq:SFactor}, the sum is over the $N_{data}$ computed properties for which agreement between the HL and pLL models is sought, $X_i^{HL}$ and $X_i^{pLL}\left(\bf{p}\right)$ are predictions of the HL and pLL models respectively, and $w_i$ sets the relative weight of the computed properties in the objective. $R\left(\bf{p}\right)$ is the regularization function described below and is included only for the training dataset. The properties that are optionally included in the objective are those listed in Section~\ref{sec:data}. 

Because only energy differences are relevant to making predictions regarding molecules, properties related to absolute energies ($E_{tot}$, $KE$, $EN_A$, $E_2$) are included in the objective of Eq.~\ref{eq:obj} in only a relative manner. This is done by taking, as a reference state for each molecule, the optimized geometry of that molecule in no external environment. The quantity $X_i$ of Eq.~\ref{eq:obj} is then the difference between the computed energy and that computed for the reference state. The objective therefore includes only the change in energy associated with distorting the molecule and placing it in an external electrostatic environment. For the remaining properties of Section~\ref{sec:data} ($E_{orb}$, and $E_{chg}$), $X_i$ of Eq.~\ref{eq:obj} is the value itself, not the difference relative to the reference state.

The weights, $w_i$, are based on the relative accuracy desired for the predicted quantities and here are set to $(\SI{2}{\kcal\per\mole})^{-1}$ for $E_{tot}$ and $(\SI{0.1}{\electronvolt})^{-1}$ for $E_{orb}$. For $E_{chg}$, the default value of 0.016 (kcal mol$^{-1}$)$^{-1}$ is sufficiently low that inclusion does not degrade performance on $E_{tot}$ or $E_{orb}$. In Section~\ref{sec:results}, the impacts of increasing the weight of $E_{chg}$ is examined. The weighting of decomposed energies is \nicefrac{1}{30}$^{th}$ that of $E_{tot}$, as discussed further in Section~\ref{sec:results}.

The parameters are optimized using the trust-region reflective algorithm~\cite{Coleman1996} as implemented within Matlab~\cite{Matlab}. To help prevent negative and overly large scaling factors from being explored, $p_0$ of Eq.~\ref{eq:SFactor} is constrained to lie between -1 and 2. The inclusion of context sensitive parameters is done in stages, such that all non-context sensitive parameters, $p_0$ of Eq.~\ref{eq:SFactor}, are first optimized, with all other parameters set to zero. As parameters associated with additional levels of context are added, all of the currently active parameters are optimized, using the values from the previous level of context as initial values.

The regularization function, $R\left(\bf{p}\right)$ of Eq.~\ref{eq:obj}, has the form
\begin{equation}\label{eq:reg}
    R\left(\bf{p}\right) = N_{data}C\|\bf{p} - \bf{p}_{init}\|
\end{equation}
where $\bf{p}_{init}$ is the initial value of the parameters for that level of context. For the initial fit of non-context sensitive parameters, $\bf{p}_{init}=0$ and thus the regularization adds a penalty for scaling factors that deviate strongly from 1. As context is added, the regularization penalizes large deviations from the scaling factors that were obtained at the previous context level. Initial studies on smaller datasets led to a value of 6 for C. 

To help prevent overtraining, decisions on when to terminate the optimization are based on performance of the model on the validation dataset. The training dataset is used to calculate the objective and its Jacobian, which the optimization algorithm then uses to generate a new trial set of parameters. If performance on the validation dataset degrades for four sequential optimization steps, this is taken as a sign of overtraining and the optimization is terminated. This process is applied at each level of context discussed above. 

The linear dependence of the scaling factors on model parameters, Eq.~\ref{eq:SFactor}, enables computational optimizations. For each molecular instance, the derivative of each operator with respect to each of the parameters is first computed. This data is then distributed such that different processor cores handle different molecular instances.  This allows the operators to be constructed quickly for arbitrary parameters. The Jacobian for the objective of Eq.~\ref{eq:obj} is obtained from finite differences, using a step size of 0.01. Analytical evaluation of the Jacobian, as obtained from the Hellmann-Feynman theorem, does not lead to as high quality in the final fits. Optimization of the parameters on the \textit{ethane} training datasets required about 100 iterations and took about 18 hours on 12 processor cores.

\section{Results} \label{sec:results}
\begin{figure}[htbp]
\centering
\includegraphics{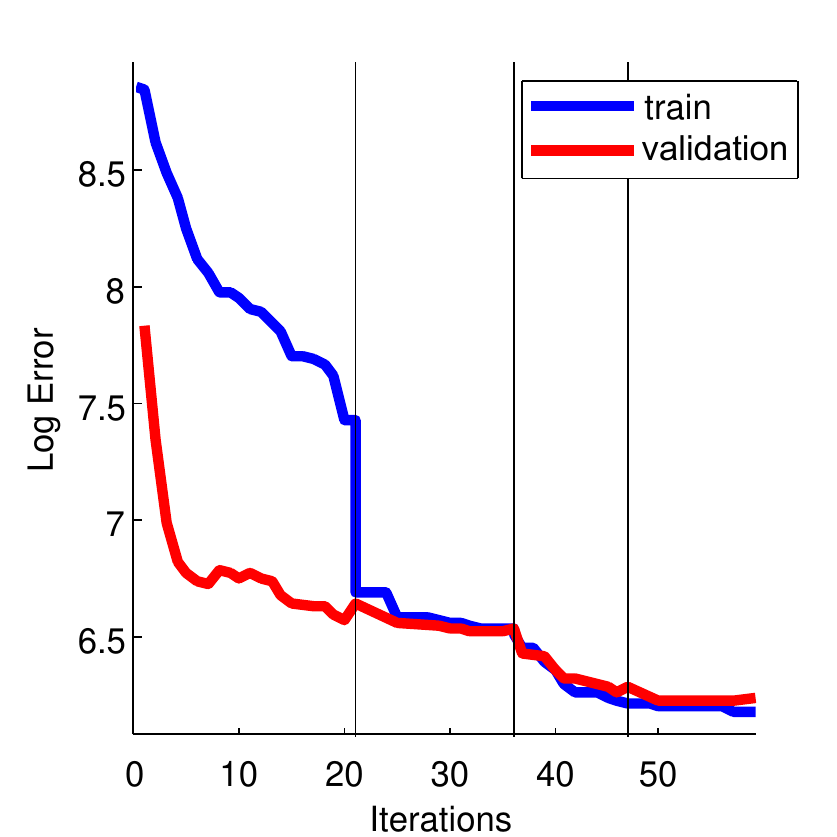}
\caption{The objective of Eq.~\ref{eq:obj}  versus iteration during training of a model on the \textit{combined} dataset. The pLL model uses the \textit{JK} policy and decomposed energies are included in the objective. Vertical lines indicate addition of the contexts of Table~\ref{tab:contexts}.}
\label{fig:iterPlot}
\end{figure}

\noindent Figure~\ref{fig:iterPlot} shows iterative improvement of the objective function of Eq.~\ref{eq:obj} during model training, with vertical lines indicating addition of context dependence to the model. The training objective includes the regularization of Eq.~\ref{eq:reg} and the strong decrease in the training objective on addition of context reflects the redefinition of the regularization in terms of deviations of the parameters from those obtained at the previous context level (Section~\ref{sec:fitting}). The greatest reduction in error occurs in the first stage, where the scaling parameters are constants, referred to as context level zero. Including the first level of context, Table~\ref{tab:contexts}, adds bond-length dependence to the parameters and also leads to a significant reduction in error. The degree to which higher levels of context lead to additional improvements is smaller and varies depending on dataset and policy.

\begin{figure*}[htbp]
\centering
\includegraphics{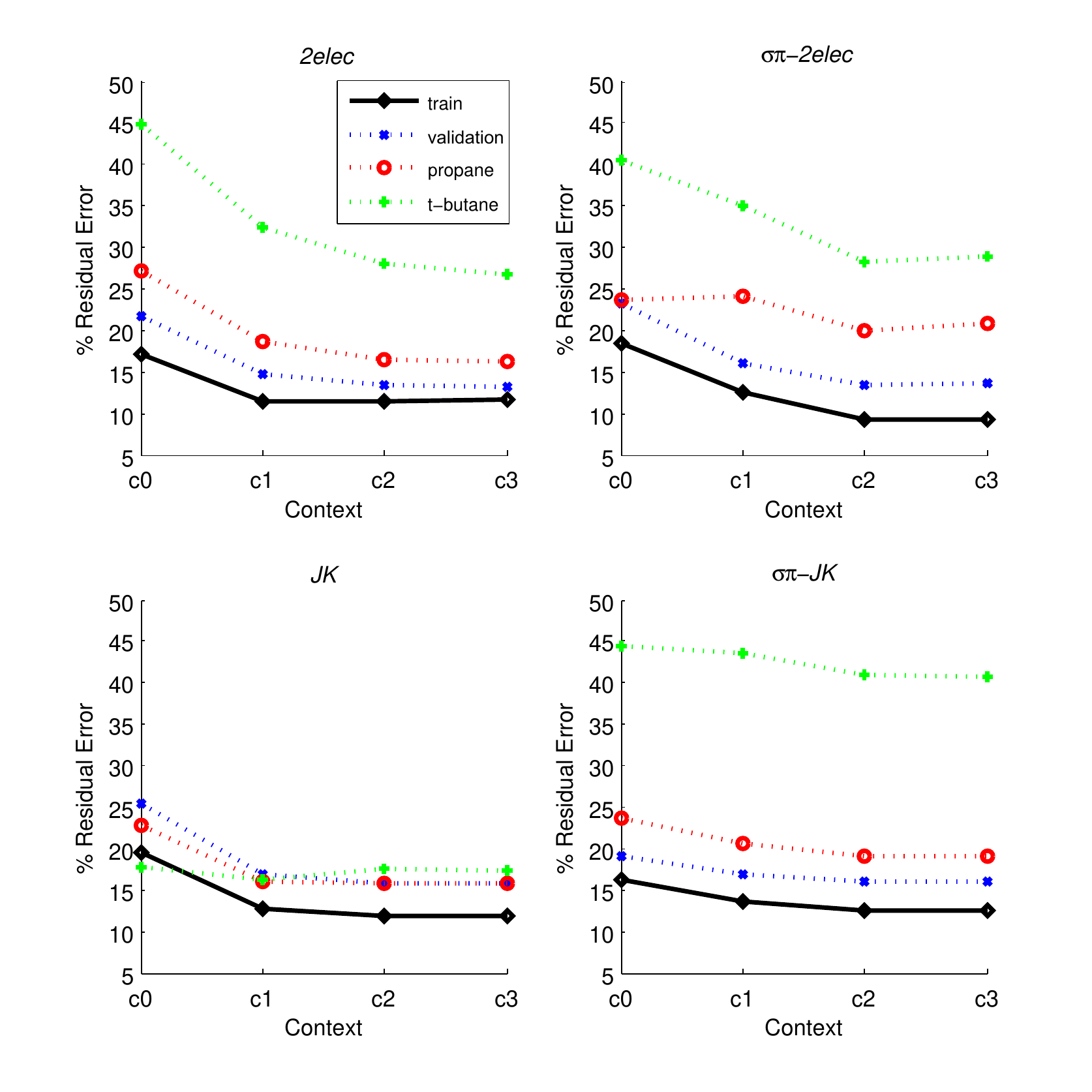}
\caption{Residual error in $E_{tot}$ as a function of context level for pLL models trained on the \textit{ethane} dataset, without inclusion of decomposed energies in the objective of Eq.~\ref{eq:obj}. Residual error is RMS disagreement between the pLL and HL models, quoted relative to initial disagreement between the LL and HL models. Panels refer to the policies of Section~\ref{sec:params}.}
\label{fig:Ethaneprop1}
\end{figure*}

\begin{figure*}[htbp]
\centering
\includegraphics{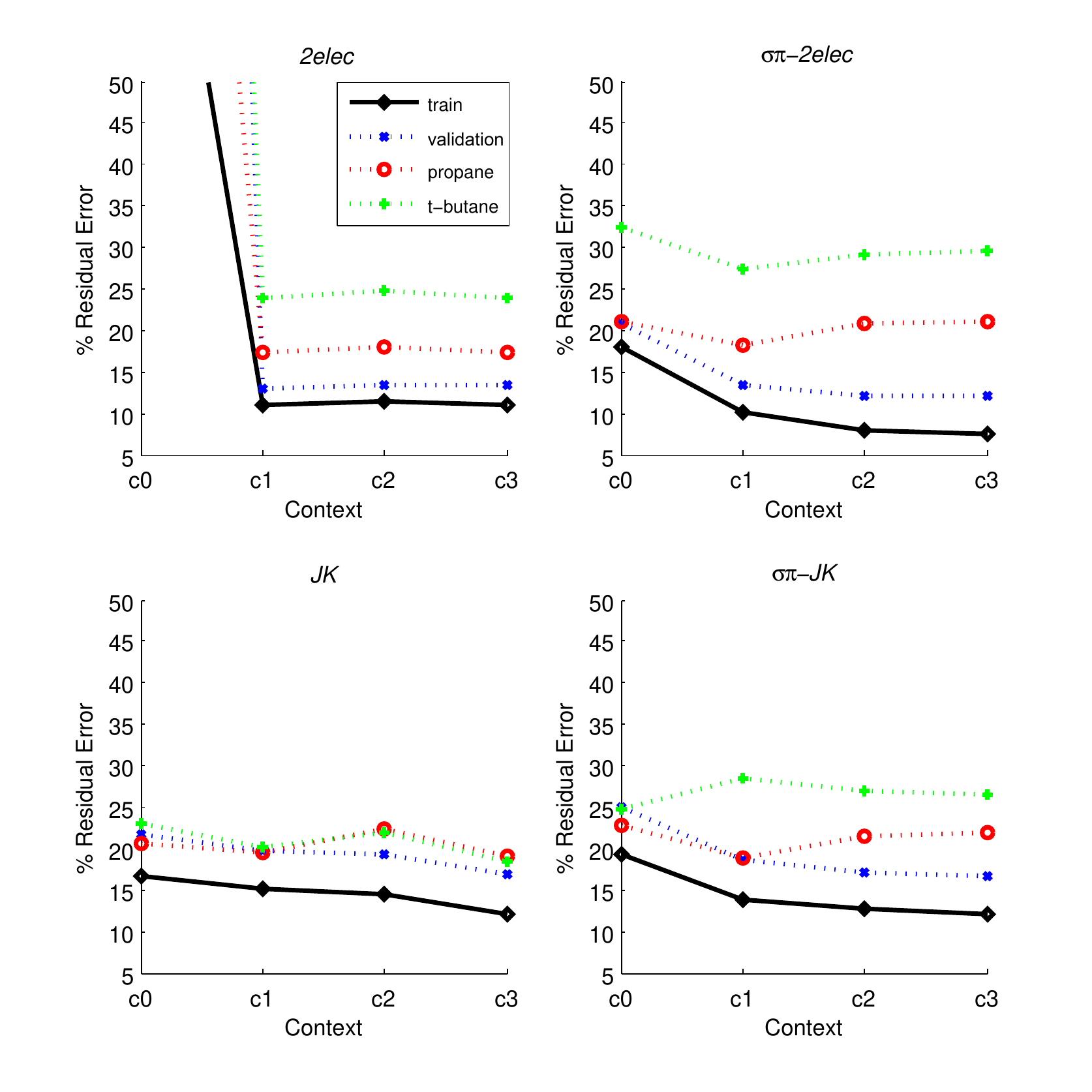}
\caption{Residual error in $E_{tot}$ as a function of context level for pLL models trained on the \textit{ethane} dataset, with inclusion of decomposed energies in the objective of Eq.~\ref{eq:obj}. Conventions are as in Fig.~\ref{fig:Ethaneprop1}.}
\label{fig:Ethaneprop2}
\end{figure*}

\begin{figure*}[htbp]
\centering
\includegraphics{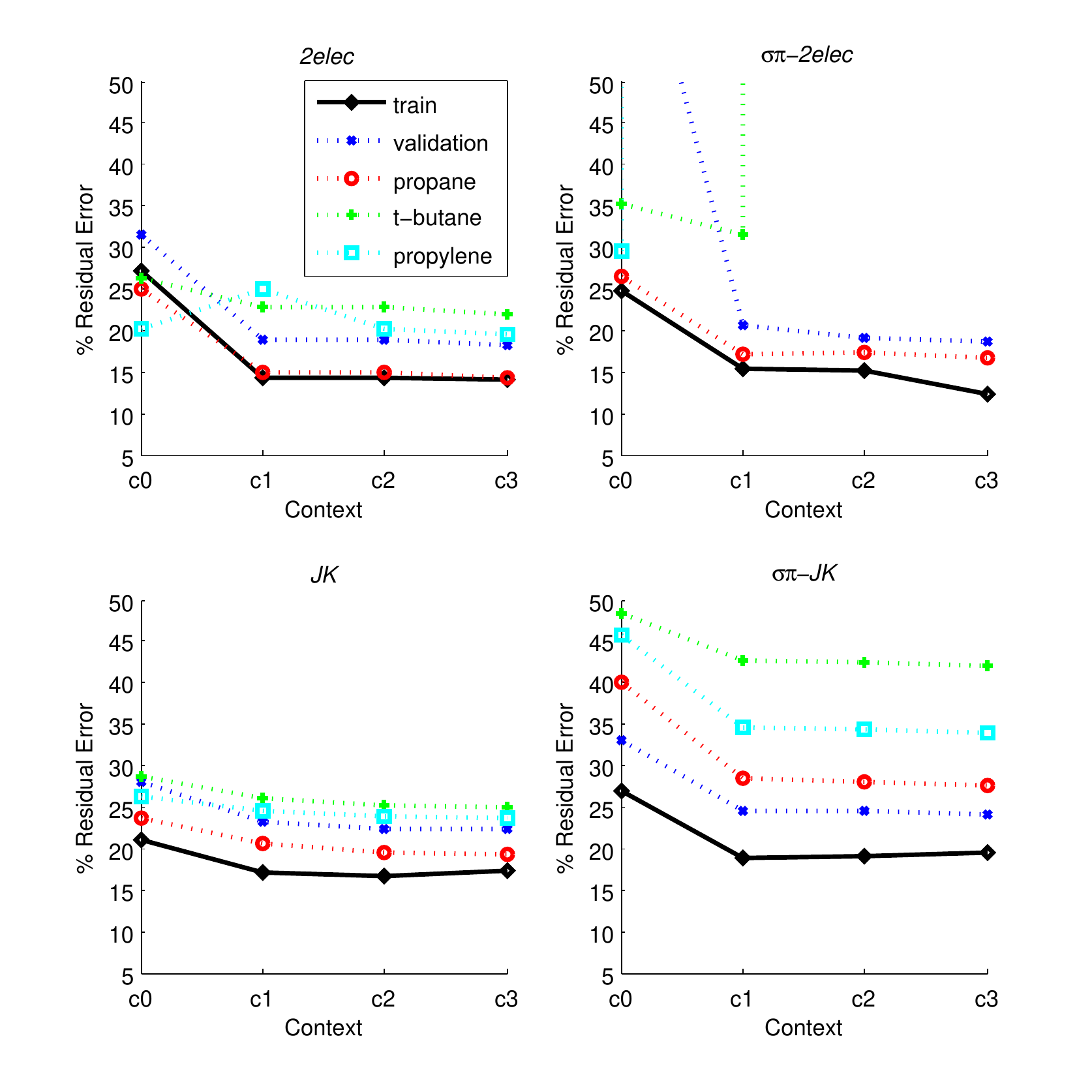}
\caption{Residual error in $E_{tot}$ as a function of context level for pLL models trained on the \textit{combined} dataset, without inclusion of decomposed energies in the objective of Eq.~\ref{eq:obj}. Conventions are as in Fig.~\ref{fig:Ethaneprop1}.}
\label{fig:EEprop1}
\end{figure*}

\begin{figure*}[htbp]
\centering
\includegraphics{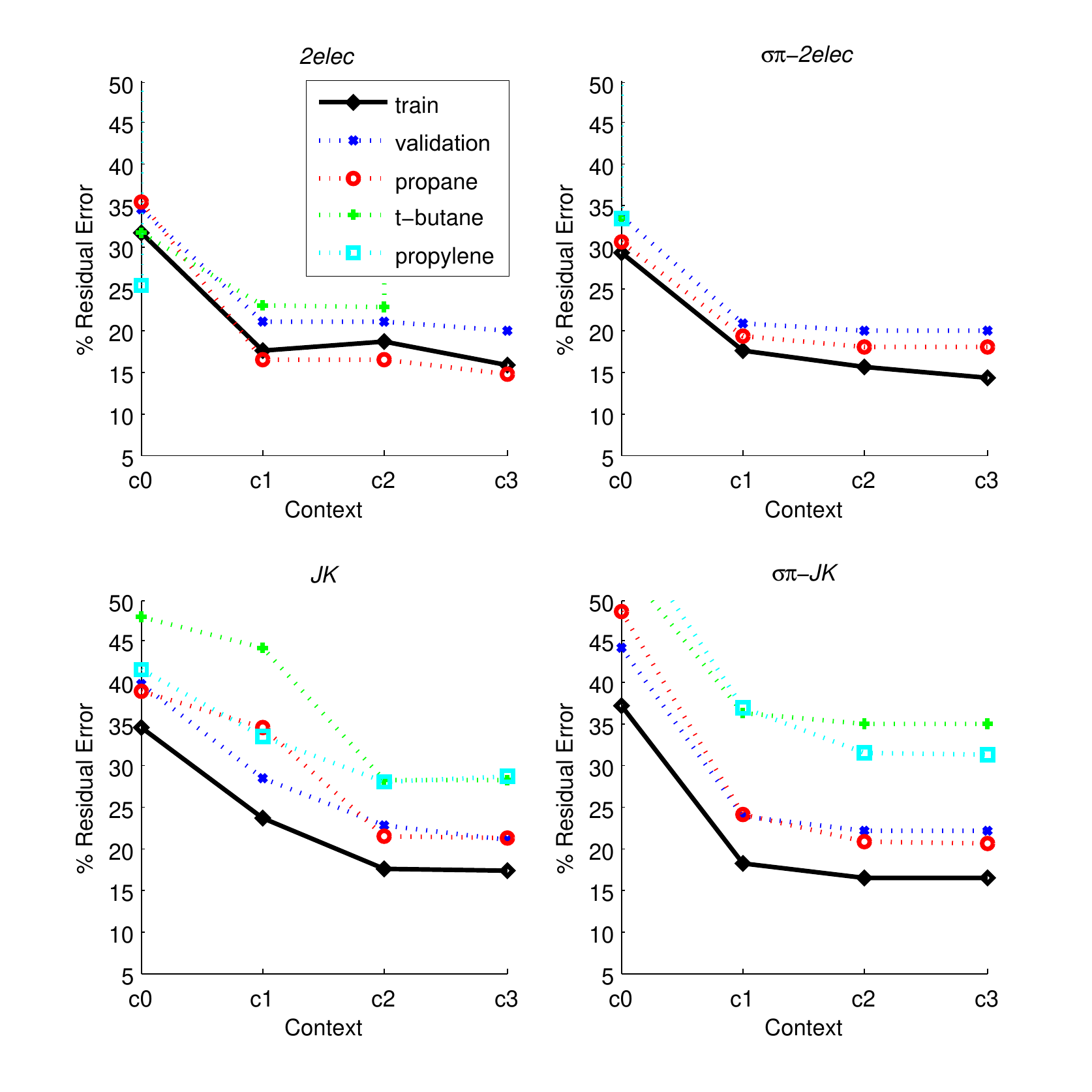}
\caption{Residual error in $E_{tot}$ as a function of context level for pLL models trained on the \textit{combined} dataset, with inclusion of decomposed energies in the objective of Eq.~\ref{eq:obj}. Conventions are as in Fig.~\ref{fig:Ethaneprop1}.}
\label{fig:EEprop2}
\end{figure*}

Models were trained on the \textit{ethane}, \textit{ethylene} and \textit{combined} datasets using the four policies discussed in Section~\ref{sec:params}. In addition, fits were done both with and without inclusion of the decomposition of the energy by operator (Section~\ref{sec:data}).  Results from each of these models are available in the \textit{Supporting Information}. Here, the focus is on models trained on the \textit{ethane} (Figures~\ref{fig:Ethaneprop1} and \ref{fig:Ethaneprop2}) and on the \textit{combined} dataset (Figures~\ref{fig:EEprop1} and \ref{fig:EEprop2} and Tables~\ref{tab:EEprop2trans} and~\ref{tab:EEprop1trans}). Models trained on ethane were not tested on ethylene or propylene since the training set did not include molecules with carbon-carbon double bonds.

\begin{table*}[htb]                                      
\centering                                               
\begin{tabular}{ccccccccccccc}                                  
\hline                                          
 &  &\multicolumn{2}{c}{Train} &\multicolumn{2}{c}{Validation} & \multicolumn{2}{c}{Propane} & \multicolumn{2}{c}{t-Butane} & \multicolumn{2}{c}{Propylene}  \\
 & Context & $n_{par}$ & E$_{tot}$ & E$_{orb}$ & E$_{tot}$ & E$_{orb}$ & E$_{tot}$ & E$_{orb}$ & E$_{tot}$ & E$_{orb}$ & E$_{tot}$ & E$_{orb}$ \\
\hline 
Initial & & & 21.47 & 4.44 & 19.21 & 4.43 & 26.68 & 5.38 & 32.69 & 5.31 & 20.03 & 2.78 \\
\hline                                          
\multirow{4}*{2elec} & c0 & 75 & 5.86 & 0.25 & 6.07 & 0.26 & 6.64 & 0.39 & 8.58 & 0.44 & 4.04 & 0.26 \\              
 & c1 & 150 & 3.05 & 0.15 & 3.63 & 0.17 & 3.96 & 0.34 & 7.44 & 0.63 & 4.99 & 0.17 \\
 & c2 & 224 & 3.08 & 0.15 & 3.62 & 0.17 & 3.97 & 0.34 & 7.48 & 0.65 & 4.05 & 0.26 \\
 & c3 & 298 & 3.00 & 0.14 & 3.52 & 0.17 & 3.84 & 0.35 & 7.18 & 0.66 & 3.90 & 0.26 \\
\hline                                                                            
\multirow{4}*{$\sigma\pi$-2elec} & c0 & 83 & 5.33 & 0.26 & 15.09 & 0.31 & 7.10 & 0.38 & 11.50 & 0.45 & 5.93 & 0.27 \\ 
 & c1 & 166 & 3.28 & 0.15 & 3.94 & 0.18 & 4.60 & 0.31 & 10.32 & 0.55 & 311.65 & 1.17 \\
 & c2 & 248 & 3.27 & 0.14 & 3.66 & 0.17 & 4.65 & 0.44 & 1514.45 & 2.28 & 740.76 & 1.74 \\
 & c3 & 330 & 2.66 & 0.13 & 3.59 & 0.16 & 4.44 & 0.50 & 2046.02 & 2.24 & 255.79 & 1.20 \\ 
\hline  
\multirow{4}*{\textit{JK}}  & c0 & 108 & 4.49 & 0.23 & 5.35 & 0.24 & 6.31 & 0.41 & 9.34 & 0.58 & 5.23 & 0.36 \\                 
 & c1 & 216 & 3.68 & 0.20 & 4.45 & 0.21 & 5.47 & 0.33 & 8.49 & 0.48 & 4.91 & 0.27 \\ 
 & c2 & 324 & 3.56 & 0.20 & 4.28 & 0.20 & 5.19 & 0.32 & 8.21 & 0.48 & 4.76 & 0.26 \\ 
 & c3 & 432 & 3.70 & 0.20 & 4.29 & 0.20 & 5.12 & 0.33 & 8.13 & 0.48 & 4.72 & 0.26 \\ 
\hline  
\multirow{4}*{$\sigma\pi$\textit{-JK}} & c0 & 125 & 5.77 & 0.24 & 6.33 & 0.26 & 10.64 & 0.40 & 15.75 & 0.44 & 9.13 & 0.20 \\     
 & c1 & 250 & 4.03 & 0.20 & 4.68 & 0.21 & 7.55 & 0.36 & 13.89 & 0.40 & 6.93 & 0.19 \\ 
 & c2 & 374 & 4.09 & 0.20 & 4.72 & 0.21 & 7.47 & 0.36 & 13.82 & 0.40 & 6.85 & 0.19 \\ 
 & c3 & 498 & 4.20 & 0.21 & 4.62 & 0.21 & 7.36 & 0.36 & 13.70 & 0.40 & 6.79 & 0.19 \\ 
\hline                              
\end{tabular}
\caption{RMS errors for models trained on the \textit{combined} dataset without inclusion of decomposed energies. $n_{par}$ is the number of parameters in the model. Large RMS errors reflect failures of SCF iterations to converge (Section~\ref{sec:results}). Units are $\SI{}{\electronvolt}$ for $E_{orb}$ and $\SI{}{\kcal\per\mole}$ for all other quantities.} 
\label{tab:EEprop1trans}  
\end{table*}

\begin{table*}[htb]                                 
\centering                
\begin{tabular}{ccccccccccccc}             
\hline
&  & \multicolumn{2}{c}{Train} & \multicolumn{2}{c}{Validation} & \multicolumn{2}{c}{Propane} & \multicolumn{2}{c}{t-Butane} & \multicolumn{2}{c}{Propylene} \\
 & Context & $n_{par}$ & E$_{tot}$ & E$_{orb}$ & E$_{tot}$ & E$_{orb}$ & E$_{tot}$ & E$_{orb}$ & E$_{tot}$ & E$_{orb}$ & E$_{tot}$ & E$_{orb}$ \\   
\hline                                            
 Initial& & & 21.47 & 4.44 & 19.21 & 4.43 & 26.68 & 5.38 & 32.69 & 5.31 & 20.03 & 2.78 \\
\hline                              
\multirow{4}*{\textit{2elec}} & c0 & 75 & 6.83 & 0.29 & 6.65 & 0.30 & 9.46 & 0.57 & 10.36 & 0.55 & 5.08 & 0.24 \\
 & c1 & 150 & 3.78 & 0.18 & 4.05 & 0.18 & 4.42 & 0.32 & 7.50 & 0.31 & 524.89 & 1.78 \\
 & c2 & 224 & 3.98 & 0.18 & 4.03 & 0.18 & 4.38 & 0.32 & 7.44 & 0.31 & 580.27 & 1.94 \\
 & c3 & 298 & 3.39 & 0.18 & 3.83 & 0.18 & 3.95 & 0.30 & 751.20 & 1.82 & 627.40 & 2.03 \\
\hline  
\multirow{4}*{$\sigma\pi$-2elec} & c0 & 83 & 6.32 & 0.29 & 6.49 & 0.29 & 8.18 & 0.53 & 10.93 & 0.49 & 6.69 & 0.27 \\                                           
 & c1 & 166 & 3.78 & 0.18 & 4.01 & 0.18 & 5.14 & 0.31 & 1105.67 & 1.41 & 851.01 & 1.60 \\ 
 & c3 & 248 & 3.36 & 0.17 & 3.82 & 0.19 & 4.82 & 0.29 & 4234.52 & 4.28 & 1743.58 & 4.06 \\
 & c3 & 330 & 3.08 & 0.17 & 3.83 & 0.18 & 4.82 & 0.31 & 4531.66 & 4.79 & 2109.05 & 4.36 \\
\hline                                                                              
\multirow{4}*{\textit{JK}} & c0 & 108 & 7.39 & 0.37 & 7.64 & 0.33 & 10.38 & 0.39 & 15.66 & 0.50 & 8.30 & 0.27 \\
 & c1 & 216 & 5.05 & 0.31 & 5.45 & 0.31 & 9.18 & 0.38 & 14.39 & 0.47 & 6.71 & 0.22 \\
 & c3 & 324 & 3.78 & 0.25 & 4.37 & 0.27 & 5.71 & 0.32 & 9.18 & 0.44 & 5.60 & 0.28 \\ 
 & c3 & 432 & 3.70 & 0.24 & 4.05 & 0.25 & 5.67 & 0.31 & 9.19 & 0.42 & 5.72 & 0.28 \\ 
\hline 
\multirow{4}*{$\sigma\pi$-\textit{JK}} & c0 & 125 & 7.99 & 0.27 & 8.45 & 0.28 & 12.92 & 0.36 & 17.86 & 0.48 & 11.85 & 0.22 \\                                            
 & c1 & 250 & 3.90 & 0.24 & 4.56 & 0.25 & 6.42 & 0.37 & 11.84 & 0.42 & 7.39 & 0.20 \\
 & c3 & 374 & 3.52 & 0.23 & 4.26 & 0.24 & 5.54 & 0.36 & 11.39 & 0.40 & 6.28 & 0.22 \\
 & c3 & 498 & 3.51 & 0.22 & 4.23 & 0.24 & 5.49 & 0.36 & 11.40 & 0.40 & 6.25 & 0.21 \\
\hline                                   
\end{tabular}             
\caption{RMS errors for models trained on the \textit{combined} dataset with inclusion of decomposed energies. Conventions are as in Table~\ref{tab:EEprop1trans}.}
\label{tab:EEprop2trans}
\end{table*} 

\begin{figure*}[htbp]
	\centering
	\includegraphics{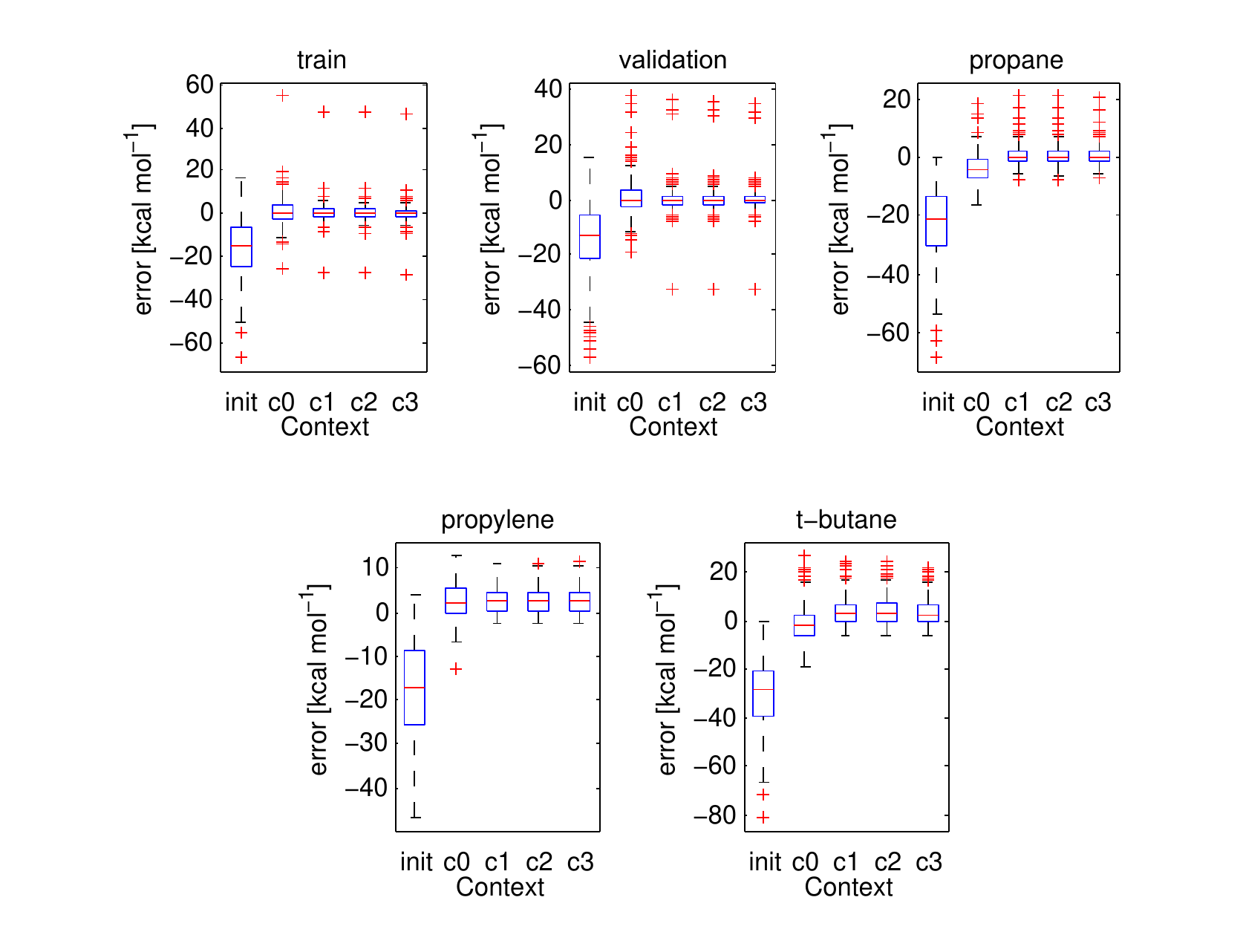}
	\caption{Boxplots of disagreements in $E_{tot}$ between HL and the best performing pLL model. The pLL model uses the \textit{2elec} policy and is trained on the \textit{combined} dataset without inclusion of decomposed energies. Initial refers to the unparametrized LL model.}
	\label{fig:boxplot}
\end{figure*}

For some sets of parameters in a pLL model, the SCF iterations may fail to lead to a converged density matrix. In such cases, predictions obtained with the unconverged density matrix are retained, and the RMS errors in $E_{tot}$ rise to hundreds of $\SI{}{\kcal\per\mole}$. During model training, inclusion of unconverged values in evaluation of the objective of Eq.~\ref{eq:obj} provides a sufficiently large penalty that the optimization algorithm rejects sets of parameters that lead to such instabilities. In reporting performance on model testing, inclusion of such unconverged values provides an indication of the degree to which a model leads to poor or unstable performance.

Three general trends emerge. First is that using separate parameters for $\sigma$ and $\pi$ interactions between \textit{p} orbitals does not improve model performance. Although the \textit{$\sigma\pi$-2elec} and \textit{$\sigma\pi$-JK} models do, in some cases, outperform their \textit{2elec} and \textit{JK} counterparts on the training dataset, performance on the test datasets is not substantially improved and in many cases degrades.  

A second general trend is that models trained on the \textit{combined} dataset typically perform better than those trained on \textit{ethane} alone. The inclusion of ethylene in the \textit{combined} dataset has the advantage of extending the training data to include a broader class of molecules. However, inclusion of ethylene also requires a single model to describe a broader class of molecules, including large amplitude rotation about a double bond. The improvement in performance seen for the \textit{combined} dataset suggests that the benefits accruing from training on more diverse data overrides the challenges associated with using a single model to describe a more diverse class of molecules. This is a promising result with regards to the ability of the approach explored here to develop models that are applicable to a diverse range of molecules. 

A third general trend is that addition of decomposed energies does little to improve model performance and often substantially degrades performance. The initial disagreement between the LL and HL models is 5 to 20 times larger for the individual energy components than for the total energy (see \textit{Supporting Information}). This indicates a large cancellation of errors occurs as the expectation values of the individual operators are summed to give the total energy. The weight of decomposed energies in Eq.~\ref{eq:obj} is set to \nicefrac{1}{30}$^{th}$ that of $E_{tot}$. With this value, the model performance on $E_{tot}$ of the training data is degraded by less than 20\%. The question is then the degree to which inclusion of this substantial amount of additional data (Table~\ref{tab:outputs}) impacts transfer to other molecules. For the \textit{ethane} dataset, comparison of Figs.~\ref{fig:Ethaneprop1} and~\ref{fig:Ethaneprop2} shows little benefit from including decomposed energies. In some cases, such as for the \textit{2elec} policy without context, inclusion of decomposed energies leads to instabilities in the resulting model. The cases where including decomposed energies improves transfer are for the $\sigma\pi$ policies. Similar results are seen for the \textit{combined} dataset in Figs.~\ref{fig:EEprop1} and~\ref{fig:EEprop2} where again, inclusion of decomposed energies improves transfer primarily for the $\sigma\pi$ policies. However, the best performing models are those with the \textit{2elec} and \textit{JK} policies, for which inclusion of decomposed energies either does little to improve or substantially lowers transfer. This can be rationalized in terms of the larger number of parameters present in the $\sigma\pi$ policies (Table~\ref{tab:EEprop1trans}). Inclusion of decomposed energies is helpful for specifying these parameters, however, better performance is obtained by using a policy with fewer parameters.

The best performing model is that obtained from the \textit{2elec} policy trained on the \textit{combined} dataset without inclusion of decomposed energies. Fig.~\ref{fig:boxplot} shows the distribution of errors in $E_{tot}$ for the training and test datasets. The distributions show that the pLL model has low systematic error such that the RMS errors reflect primarily the width of the distribution. Outliers are also present for both the initial LL model and the trained pLL models. Removal of these outliers would lower the reported RMS errors by approximately 0.3 to $\SI{0.5}{\kcal\per\mole}$, for the well-performing models.
\begin{table*}[htb]
\centering
\resizebox{\textwidth}{!}{\begin{tabular}{ccccccccccccccccc}
\hline
 Dataset & Weight &\multicolumn{3}{c}{Train}&\multicolumn{3}{c}{Validation}&\multicolumn{3}{c}{Propane} &\multicolumn{3}{c}{tButane} &\multicolumn{3}{c}{propylene}\\
 &$(\SI{}{\kcal\per\mole})^{-1}$& E$_{tot}$ & E$_{chg}$ & E$_{orb}$ & E$_{tot}$ & E$_{chg}$ & E$_{orb}$ & E$_{tot}$ & E$_{chg}$ & E$_{orb}$ & E$_{tot}$ & E$_{chg}$ & E$_{orb}$ & E$_{tot}$ & E$_{chg}$ & E$_{orb}$ \\
\hline
  \multirow{6}*{\textit{ethane}} & 0.016 & 2.55 & 4.15 & 0.18 & 2.57 & 4.14 & 0.23 & 4.31 & 5.90 & 0.27 & 8.72 & 7.32 & 0.37 & - & - & - \\
  & 0.16 & 2.42 & 3.98 & 0.18 & 2.48 & 3.98 & 0.23 & 4.32 & 5.74 & 0.26 & 8.58 & 7.13 & 0.35 & - & - & -  \\
  & 0.20 & 2.19 & 3.50 & 0.16 & 2.24 & 3.49 & 0.20 & 3.99 & 5.06 & 0.22 & 7.58 & 6.33 & 0.30 & - & - & - \\
  & 0.27 & 2.53 & 3.31 & 0.16 & 2.51 & 3.31 & 0.20 & 4.05 & 4.74 & 0.22 & 7.45 & 5.98 & 0.30 & - & - & - \\
  & 0.40 & 2.71 & 3.09 & 0.17 & 2.93 & 3.10 & 0.19 & 4.18 & 4.32 & 0.25 & 7.22 & 5.51 & 0.33 & - & - & - \\
  & 0.80 & 2.81 & 2.98 & 0.20 & 3.40 & 3.01 & 0.22 & 4.86 & 4.07 & 0.29 & 7.02 & 5.17 & 0.36 & - & - & - \\
  \hline
  \multirow{6}*{\textit{combined}} & 0.016 & 3.00 & 4.27 & 0.14 & 3.52 & 4.51 & 0.17 & 3.84 & 4.86 & 0.35 & 7.18 & 6.11 & 0.66 & 3.91 & 5.46 & 0.26 \\
  & 0.16 & 3.09 & 4.25 & 0.15 & 3.65 & 4.46 & 0.17 & 3.92 & 4.80 & 0.44 &  7.36 & 6.03 & 0.76 & 4.15 & 5.40 & 0.31 \\
  & 0.20 & 3.78 & 3.53 & 0.15 & 4.14 & 3.72 & 0.15 & 5.15 & 4.57 & 0.45 & 10.20 & 5.83 & 0.88 & 5.43 & 4.93 & 0.34 \\
  & 0.27 & 4.00 & 3.38 & 0.17 & 4.45 & 3.56 & 0.18 & 5.49 & 4.57 & 0.51 & 11.59 & 5.89 & 1.17 & 22.60 & 5.11 & 0.56 \\
  & 0.40 & 3.99 & 3.36 & 0.16 & 4.47 & 3.54 & 0.18 & 5.59 & 4.57 & 0.52 & 11.88 & 5.91 & 1.19 & 24.84 & 5.11 & 0.58 \\
  & 0.80 & 4.00 & 3.28 & 0.17 & 4.53 & 3.46 & 0.19 & 6.09 & 4.59 & 0.57 & 13.99 & 6.02 & 1.39 & 89.12 & 5.51 & 1.04 \\
\hline
\end{tabular}}
\caption{Effects of increasing the weighting of $E_{chg}$ in the objective of Eq.~\ref{eq:obj}, for pLL models using the \textit{2elec} policy with full context trained without inclusion of decomposed energies.}
\label{tab:EextW}
\end{table*}

The effects of increasing the weighting of $E_{chg}$ in the objective of Eq.~\ref{eq:obj} is examined in Table~\ref{tab:EextW}, for the best performing model. For training on the \textit{combined} dataset, increasing the weight by an order of magnitude from the default value of 0.016 $(\SI{}{\kcal\per\mole})^{-1}$ leads to an improved description of the charge distribution, as measured by $E_{chg}$, at the expense of $E_{tot}$. According to the arguments accompanying Eq.~\ref{eq:rhorhobasis}, an improved description of the charge distribution should lead to improvements in the longer-range interactions. However, the performance of the model on the larger test molecules degrades with increased weighting of $E_{chg}$. This suggests the longer-range interactions are not the primary source of error in those systems. A somewhat different behavior is observed for the smaller \textit{ethane} dataset. Here, as the  the weighting of $E_{chg}$ is increased, the training error in $E_{tot}$ first decreases and then increases. This suggests that additional information regarding interaction of the charge density with the external charges aids the training process at intermediate weights and, only at large weights, does improved performance on $E_{chg}$ come at the expense of degraded performance on $E_{tot}$. Similar behavior is seen for the validation dataset, which includes  different geometries of the training molecules. For the larger test molecules, increasing the weight of $E_{chg}$ improves the performance on $E_{tot}$. These observations can be rationalized by viewing the interaction with external charges as simply additional data. For the smaller \textit{ethane} dataset, this additional data aids the training. For the larger \textit{combined} dataset, placing increased weight on this additional data degrades performance. 

\FloatBarrier

\section{Discussion}\label{sec:discussion}
\noindent The goal of this work is to develop flexible and systematically-improvable means to take advantage of molecular similarity in quantum chemical computations. The approach explored here embeds parameters in a LL \textit{ab initio} Hamiltonian and adjusts these to obtain agreement with predictions of a HL \textit{ab initio} Hamiltonian. This approach bridges between the flexible models of machine learning, such as neural nets, and the model Hamiltonians of SEQC. Model forms based on quantum chemical Hamiltonians may have advantages that stem from being more closely related to chemical phenomena than generic forms such as neural nets. For example, models trained on small molecules may incorporate sufficient information regarding molecular fragments as to be applicable to larger systems. Here, models trained on ethane and ethylene transfer reasonably well to propane, propylene and butane. In addition, models trained on ethane and ethylene perform better than models trained on ethane alone. This indicates that the benefits gained from including the additional ethylene data in the training override the challenges associated with using a single model to describe a more diverse class of molecules. 

The general approach of embedding parameters in a LL \textit{ab initio} Hamiltonian provides a flexible approach to model construction. Considerable flexibility may stem from using different LL Hamiltonians. However, the current work considers only different schemes, or policies, for embedding scaling factors in a minimal basis \textit{ab initio} Hamiltonian. The policies differ along two dimensions. One dimension is whether different scaling factors are used for $\sigma$ versus $\pi$ interactions between $p$ orbitals, indicated as $\sigma\pi$ policies. For the datasets considered here, better performance is obtained when $\sigma$ versus $\pi$ interactions were not treated separated. The other dimension along which the policies differ is whether scaling factors were applied directly to the two-electron matrix elements, \textit{2elec}, or to the matrix elements of the $J$ and $K$ operators in the Fock matrix, \textit{JK}. The \textit{2elec} policy led to somewhat better performance than the \textit{JK} policy. That the best performing policy, \textit{2elec}, is the policy with the fewest parameters may indicate that the parameters in the other policies are underdetermined (Table~\ref{tab:EEprop1trans}). If this is the case, the other policies may perform better on datasets involving larger classes of molecules. 

Another means through which the model is made flexible and improvable is by making the scaling factors functions of context variables that capture the environment of the molecular fragment. Here, the model is first trained with no context dependence. A first level of context is added by making the scaling factors linear functions of bond lengths. This leads to substantial performance enhancements. Two additional levels of context are added that relate to the electronic structure, via the charge and bond order predicted by the unparametrized LL model. Depending on the policy and training dataset, these additional levels can lead to additional small improvements. Context variables are a means to add considerable flexibility to the model, but taking better advantage of this flexibility may require discovery of variables that better describe the molecular context. 

Extending the types of data used to train the model is a potential means to improve model performance. One source of additional data is the interaction with external charges. This provides information on the electron density that is comparable across LL and HL models. Inclusion of this additional data leads to small improvements for the \textit{ethane} dataset but not for the \textit{combined} dataset. Another source of additional data is decomposition of the energy by operator. Inclusion of such data improves performance for $\sigma\pi$ policies in some cases, but tends to degrade performance for the other policies. This is consistent with the above conclusion that the larger number of parameters in the $\sigma\pi$ policies leads to an underdetermined model. The improvement in the $\sigma\pi$ policies resulting from inclusion of decomposed energies is, however, not sufficient to make them competitive with the other policies. Decomposition of the energy by operator is therefore not found to lead to benefits in model training.  

Decomposition of the energy by operator has the advantage of being uniquely defined, such that it can be applied unambiguously in both the LL and HL models.  Decomposition of the energy by molecular fragment may provide information more relevant to the training, since the embedded parameters are associated with local interactions on and between bonded atoms. However, such decompositions are not unique and require division of the electron density in a manner that is compatible across the LL and HL models\cite{quambo,bader1991}.  

A means through which future work may increase the flexibility of the pLL models is by relaxing some constraints present in the current approach to applying scaling factors. In particular, the \textit{p} orbitals on an atom are currently treated as equivalent such that, for example, a single scaling factor is used for the \textit{p} subblock of the kinetic energy operator on a heavy atom. When the Quambo method is used to transform the high-level electron densities to a minimal basis form, the resulting \textit{p} orbitals do not retain this equivalency\cite{quambo}. 

Addition of parametrized core-core potentials is an additional avenue through which the model can be made more flexible. The current pLL models do not include such terms because the core electrons are included explicitly using basis functions that are identical to those in the HL model. In SEQC and DFTB, core-core terms substantially enhance the accuracy.  A possible approach is to follow the above training of the pLL model with an additional stage that trains only core-core terms to reproduce HL molecular forces. This would be analogous to the DFTB approach, which first derives the electronic Hamiltonian and then fits the core-core potentials\cite{Elstner1998,Cui2001,Elstner2006,Gaus2012}. The distinction being that here both the electronic Hamiltonian and the core-core potentials would be obtained empirically.  

In the current work, the parameters in the pLL model were used only to compensate for errors arising from the use of a minimal basis set in the LL model. The ability of this approach to compensate for errors arising from the absence of electron correlation in a LL model is yet to be explored. The ability of parametrized LL models to reduce errors in the original LL model by over 75\%, for molecules larger than those included in the training data, suggests this approach has promise for using molecular similarity to reduce the computational cost of quantum chemical computations.  

\section*{Acknowledgements} 
Supported by the National Science Foundation 1027985 and 1135553.

\bibliography{msqc}

\newpage 
\section{Supporting information}

Models were trained on the \textit{ethane}, \textit{ethylene} and \textit{combined} datasets using the four policies discussed in Section~III of the main text. In addition, fits were done both with and without inclusion of the decomposition of the energy by operator (see Section~II of main text).  Subsets of these results are presented and discussed in Section VI of the main text. Tables~\ref{tab:Ethaneprop2} through~\ref{tab:EEprop1transSP} list results from all fits. Table~\ref{tab:Ethaneprop2},~\ref{tab:Ethylprop2}, and~\ref{tab:EEprop2} report values for the energy decomposed by operator. The RMS errors in energy components are substantially larger than in total energy. This indicates a large cancellation of error between the components when they are summed to the total energy.

\begin{table}[h]   
\centering                                                                
\begin{tabular}{cccccccccccccc}              
\hline                                            
 &  & \multicolumn{2}{c}{KE} & \multicolumn{2}{c}{EN$_H$} & \multicolumn{2}{c}{EN$_A$} & \multicolumn{2}{c}{E2} & \multicolumn{2}{c}{E$_{tot}$} & \multicolumn{2}{c}{E$_{orb}$} \\
 &  & train & validation & train & validation & train & validation & train & validation & train & validation & train & validation \\
\hline  
 & Initial & 184.80 & 165.05 & 29.97 & 28.04 & 380.71 & 334.75 & 147.49 & 132.57 & 21.98 & 19.44 & 5.57 & 5.57 \\                                                                 
\hline  
\multirow{4}*{2elec} & c0 & 124.36 & 102.66 & 24.55 & 24.17 & 339.54 & 326.79 & 113.80 & 282.75 & 21.56 & 81.86 & 0.61 & 0.74 \\                                                  
 & c1 & 33.84 & 34.28 & 17.61 & 16.61 & 105.33 & 109.99 & 52.65 & 54.63 & 2.42 & 2.50 & 0.18 & 0.21 \\
 & c2 & 32.60 & 33.22 & 16.36 & 15.40 & 101.40 & 106.28 & 49.61 & 51.99 & 2.49 & 2.60 & 0.17 & 0.20 \\
 & c3 & 30.49 & 31.34 & 14.69 & 13.80 & 95.78 & 101.43 & 46.13 & 48.54 & 2.44 & 2.59 & 0.16 & 0.19 \\ 
\hline 
\multirow{4}*{$\sigma\pi-2elec$} & c0 & 83.42 & 86.07 & 24.97 & 23.51 & 153.44 & 150.39 & 55.88 & 56.81 & 3.95 & 4.04 & 0.28 & 0.26 \\                                            
 & c1 & 29.03 & 30.32 & 13.63 & 12.49 & 95.74 & 93.51 & 46.12 & 43.31 & 2.24 & 2.61 & 0.14 & 0.17 \\   
 & c2 & 26.48 & 27.27 & 10.72 & 9.71 & 78.99 & 84.80 & 38.45 & 37.68 & 1.76 & 2.37 & 0.13 & 0.16 \\    
 & c3 & 26.19 & 26.73 & 10.34 & 9.37 & 76.68 & 82.52 & 37.24 & 36.72 & 1.66 & 2.36 & 0.13 & 0.16 \\    
\hline 
\multirow{4}*{JK} & c0 & 81.24 & 83.91 & 31.91 & 29.74 & 202.60 & 198.78 & 59.84 & 59.90 & 3.68 & 4.22 & 0.25 & 0.23 \\                                                           
 & c1 & 56.03 & 56.90 & 29.27 & 27.25 & 174.75 & 170.49 & 51.67 & 53.18 & 3.31 & 3.81 & 0.23 & 0.22 \\ 
 & c2 & 47.70 & 48.09 & 24.52 & 22.79 & 146.62 & 145.34 & 46.26 & 48.79 & 3.19 & 3.76 & 0.21 & 0.21 \\ 
 & c3 & 48.76 & 49.38 & 23.64 & 21.97 & 141.53 & 141.32 & 45.28 & 48.27 & 2.66 & 3.29 & 0.20 & 0.20 \\ 
\hline 
\multirow{4}*{$\sigma\pi$-JK} & c0 & 86.26 & 89.40 & 32.72 & 30.58 & 198.50 & 197.01 & 56.21 & 54.38 & 4.24 & 4.85 & 0.24 & 0.24 \\
 & c1 & 63.76 & 64.81 & 29.26 & 27.28 & 173.81 & 170.00 & 51.55 & 50.53 & 3.05 & 3.63 & 0.23 & 0.22 \\
 & c2 & 51.16 & 52.05 & 24.22 & 22.55 & 155.61 & 151.03 & 47.67 & 46.96 & 2.80 & 3.32 & 0.21 & 0.20 \\
 & c3 & 46.08 & 46.48 & 21.44 & 19.94 & 140.29 & 136.98 & 45.13 & 44.77 & 2.65 & 3.22 & 0.20 & 0.19 \\
\hline 
\end{tabular}                                                    
\caption{RMS errors for models trained on the \textit{ethane} dataset with inclusion of decomposed energies. Units are $\SI{}{\electronvolt}$ for $E_{orb}$ and $\SI{}{\kcal\per\mole}$ for all other quantities.} 
\label{tab:Ethaneprop2}
\end{table} 

\begin{table}[h]
\centering
\begin{tabular}{cccccccccc}
\hline 
 &  & \multicolumn{2}{c}{Train} & \multicolumn{2}{c}{Validation} & \multicolumn{2}{c}{Propane} & \multicolumn{2}{c}{t-Butane} \\
 & Context & E$_{tot}$ & E$_{orb}$ & E$_{tot}$ & E$_{orb}$ & E$_{tot}$ & E$_{orb}$ & E$_{tot}$ & E$_{orb}$ \\
\hline 
 & Initial & 21.98 & 5.57 & 19.44 & 5.57 & 26.68 & 5.38 & 32.69 & 5.31 \\
\hline 
\multirow{4}*{2elec} & c0 & 21.56 & 0.61 & 81.86 & 0.74 & 49.63 & 0.81 & 127.39 & 0.73 \\
 & c1 & 2.42 & 0.18 & 2.50 & 0.21 & 4.64 & 0.30 & 7.81 & 0.43 \\
 & c2 & 2.49 & 0.17 & 2.60 & 0.20 & 4.81 & 0.32 & 8.12 & 0.48 \\
 & c3 & 2.44 & 0.16 & 2.59 & 0.19 & 4.62 & 0.32 & 7.80 & 0.55 \\
\hline      
\multirow{4}*{$\sigma\pi$-2elec} & c0 & 3.95 & 0.28 & 4.04 & 0.26 & 5.64 & 0.38 & 10.57 & 0.33 \\
 & c1 & 2.24 & 0.14 & 2.61 & 0.17 & 4.86 & 0.32 & 8.94 & 0.32 \\
 & c3 & 1.76 & 0.13 & 2.37 & 0.16 & 5.57 & 0.30 & 9.51 & 0.32 \\
 & c3 & 1.66 & 0.13 & 2.36 & 0.16 & 5.64 & 0.29 & 9.68 & 0.31 \\
\hline 
\multirow{4}*{JK} & c0 & 3.68 & 0.25 & 4.22 & 0.23 & 5.49 & 0.50 & 7.53 & 0.65 \\
 & c1 & 3.31 & 0.23 & 3.81 & 0.22 & 5.21 & 0.45 & 6.61 & 0.43 \\
 & c3 & 3.19 & 0.21 & 3.76 & 0.21 & 5.94 & 0.42 & 7.17 & 0.41 \\
 & c3 & 2.66 & 0.20 & 3.29 & 0.20 & 5.07 & 0.41 & 6.04 & 0.43 \\
\hline  
\multirow{4}*{$\sigma\pi$-JK} & c0 & 4.24 & 0.24 & 4.85 & 0.24 & 6.04 & 0.59 & 8.12 & 0.70 \\
 & c1 & 3.05 & 0.23 & 3.63 & 0.22 & 5.00 & 0.54 & 9.32 & 0.60 \\
 & c3 & 2.80 & 0.21 & 3.32 & 0.20 & 5.72 & 0.52 & 8.78 & 0.62 \\
 & c3 & 2.65 & 0.20 & 3.22 & 0.19 & 5.83 & 0.50 & 8.70 & 0.61 \\
\hline
\end{tabular}
\caption{RMS errors for models trained on the \textit{ethane} dataset with inclusion of decomposed energies. Units are $\SI{}{\electronvolt}$ for $E_{orb}$ and $\SI{}{\kcal\per\mole}$ for all other quantities.}
\label{tab:Ethaneprop2trans}
\end{table}

\begin{table}[h] 
\centering        
\begin{tabular}{cccccccccc}
\hline                     
 &  & \multicolumn{2}{c}{Train} & \multicolumn{2}{c}{Validation} & \multicolumn{2}{c}{Propane} & \multicolumn{2}{c}{t-Butane} \\
 & Context  & E$_{tot}$ & E$_{orb}$ & E$_{tot}$ & E$_{orb}$ & E$_{tot}$ & E$_{orb}$ & E$_{tot}$ & E$_{orb}$ \\                    
\hline            
Initial &  & 21.98 & 5.57 & 19.44 & 5.57 & 26.68 & 5.38 & 32.69 & 5.31 \\
\hline            
\multirow{4}*{2elec} & c0 & 3.75 & 0.29 & 4.24 & 0.35 & 7.28 & 0.70 & 14.70 & 1.47 \\
 & c1 & 2.49 & 0.21 & 2.86 & 0.27 & 4.96 & 0.32 & 10.62 & 0.46 \\
 & c2 & 2.52 & 0.19 & 2.62 & 0.25 & 4.40 & 0.29 & 9.19 & 0.40 \\ 
 & c3 & 2.55 & 0.18 & 2.57 & 0.23 & 4.31 & 0.27 & 8.72 & 0.36 \\ 
\hline                  
\multirow{4}*{$\sigma\pi$-2elec} & c0 & 4.05 & 0.26 & 4.51 & 0.26 & 6.32 & 0.33 & 13.20 & 0.52 \\
 & c1 & 2.75 & 0.17 & 3.13 & 0.18 & 6.46 & 0.35 & 11.45 & 0.85 \\
 & c3 & 2.06 & 0.14 & 2.61 & 0.16 & 5.35 & 0.28 & 9.27 & 0.45 \\
 & c3 & 2.05 & 0.13 & 2.63 & 0.16 & 5.57 & 0.28 & 9.44 & 0.43 \\
\hline 
\multirow{4}*{JK} & c0 & 4.29 & 0.19 & 4.93 & 0.19 & 6.07 & 0.45 & 5.81 & 0.68 \\
 & c1 & 2.81 & 0.17 & 3.26 & 0.18 & 4.25 & 0.42 & 5.34 & 0.63 \\
 & c3 & 2.62 & 0.17 & 3.09 & 0.18 & 4.20 & 0.42 & 5.71 & 0.63 \\
 & c3 & 2.61 & 0.17 & 3.08 & 0.18 & 4.19 & 0.42 & 5.69 & 0.63 \\
\hline            
\multirow{4}*{$\sigma\pi$-JK} & c0 & 3.56 & 0.21 & 3.71 & 0.22 & 6.28 & 0.55 & 14.50 & 0.66 \\
 & c1 & 3.01 & 0.18 & 3.27 & 0.20 & 5.47 & 0.49 & 14.23 & 0.60 \\
 & c3 & 2.77 & 0.17 & 3.09 & 0.20 & 5.11 & 0.47 & 13.38 & 0.58 \\
 & c3 & 2.77 & 0.17 & 3.09 & 0.19 & 5.08 & 0.48 & 13.28 & 0.59 \\
\hline      
\end{tabular}
\caption{RMS errors for models trained on the \textit{ethane} dataset without inclusion of decomposed energies. Units are $\SI{}{\electronvolt}$ for $E_{orb}$ and $\SI{}{\kcal\per\mole}$ for $E_{tot}$.}
\label{tab:Ethaneprop1trans}
\end{table} 

\begin{table}[h]                                                         
\centering 
\begin{tabular}{cccccccccccccc}
\hline 
 &  & \multicolumn{2}{c}{KE} & \multicolumn{2}{c}{EN$_H$} & \multicolumn{2}{c}{EN$_A$} & \multicolumn{2}{c}{E2} & \multicolumn{2}{c}{E$_{tot}$} & \multicolumn{2}{c}{E$_{orb}$} \\
 & Context & train & validation & train & validation & train & validation & train & validation & train & validation & train & validation \\
\hline 
Initial &  & 147.15 & 150.60 & 24.65 & 24.43 & 424.78 & 434.49 & 133.36 & 140.54 & 20.95 & 18.98 & 2.88 & 2.87 \\                                                                 
\hline  
\multirow{4}*{2elec} & c0 & 68.33 & 78.27 & 25.53 & 27.01 & 258.20 & 258.19 & 70.30 & 92.05 & 5.51 & 7.08 & 0.26 & 0.30 \\ 
 & c1 & 56.68 & 57.68 & 20.52 & 20.08 & 206.49 & 210.16 & 54.77 & 64.58 & 3.95 & 5.17 & 0.19 & 0.21 \\ 
 & c2 & 54.23 & 109.45 & 17.93 & 18.37 & 196.03 & 235.24 & 50.82 & 93.42 & 3.86 & 10.43 & 0.18 & 0.29 \\
 & c3 & 53.16 & 148.63 & 16.85 & 18.07 & 190.07 & 260.91 & 48.08 & 118.78 & 3.86 & 21.05 & 0.18 & 0.37 \\  
\hline          
\multirow{4}*{$\sigma\pi$-2elec} & c0 & 96.36 & 103.33 & 23.91 & 22.19 & 430.88 & 422.60 & 87.14 & 80.52 & 14.20 & 11.46 & 0.36 & 0.38 \\                                         
 & c1 & 83.91 & 100.66 & 24.17 & 25.26 & 392.64 & 399.20 & 78.38 & 128.92 & 10.67 & 14.68 & 0.32 & 0.36 \\
 & c2 & 74.08 & 73.16 & 29.09 & 23.82 & 324.21 & 311.16 & 142.18 & 72.99 & 19.19 & 8.60 & 0.34 & 0.28 \\
 & c3 & 64.89 & 68.27 & 27.08 & 23.61 & 301.80 & 290.42 & 106.63 & 68.65 & 17.85 & 8.11 & 0.30 & 0.27 \\
\hline      
\multirow{4}*{JK} & c0 & 56.83 & 63.71 & 25.47 & 25.81 & 272.45 & 247.86 & 59.46 & 66.94 & 4.61 & 5.60 & 0.21 & 0.23 \\ 
 & c1 & 54.12 & 60.51 & 24.62 & 25.03 & 259.70 & 241.57 & 58.35 & 65.14 & 4.43 & 5.47 & 0.21 & 0.23 \\ 
 & c2 & 49.90 & 55.51 & 23.46 & 23.63 & 263.70 & 229.72 & 56.58 & 63.75 & 4.31 & 5.18 & 0.21 & 0.22 \\ 
 & c3 & 49.66 & 55.30 & 23.23 & 23.65 & 243.33 & 228.40 & 56.56 & 63.30 & 4.15 & 5.21 & 0.20 & 0.22 \\ 
\hline 
\multirow{4}*{$\sigma\pi$-JK} & c0 & 68.06 & 77.78 & 28.89 & 28.97 & 264.88 & 239.65 & 62.73 & 70.21 & 5.58 & 6.10 & 0.23 & 0.25 \\                                               
 & c1 & 57.21 & 65.24 & 26.14 & 26.14 & 240.27 & 222.16 & 56.74 & 62.37 & 4.50 & 5.29 & 0.21 & 0.22 \\
 & c2 & 56.87 & 64.07 & 25.44 & 25.35 & 239.93 & 219.75 & 57.09 & 62.59 & 7.04 & 5.29 & 0.23 & 0.22 \\
 & c3 & 55.46 & 62.58 & 24.99 & 24.88 & 245.44 & 215.11 & 54.69 & 61.14 & 4.25 & 5.17 & 0.20 & 0.22 \\
\hline                                                             
\end{tabular}
\caption{RMS errors for models trained on the \textit{ethylene} dataset with inclusion of decomposed energies. Units are $\SI{}{\electronvolt}$ for $E_{orb}$ and $\SI{}{\kcal\per\mole}$ for all other quantities.}
\label{tab:Ethylprop2} 
\end{table} 

\begin{table}[h]                                                                            
\centering                                                                                   
\begin{tabular}{cccccccc}                                                                    
\hline                                                                                       
 &  & \multicolumn{2}{c}{Train} & \multicolumn{2}{c}{Validation} & \multicolumn{2}{c}{Propylene} \\
 &  & E$_{tot}$ & E$_{orb}$ & E$_{tot}$ & E$_{orb}$ & E$_{tot}$ & E$_{orb}$ \\               
\hline                                                                                       
 & Initial & 20.95 & 2.88 & 18.98 & 2.87 & 20.03 & 2.78 \\                                   
\hline                                                                                       
\multirow{4}*{2elec} & c0 & 5.51 & 0.26 & 7.08 & 0.30 & 29.17 & 0.47 \\                      
 & c1 & 3.95 & 0.19 & 5.17 & 0.21 & 4.84 & 0.38 \\                                           
 & c2 & 3.86 & 0.18 & 10.43 & 0.29 & 5.00 & 0.38 \\                                          
 & c3 & 3.86 & 0.18 & 21.05 & 0.37 & 40.71 & 0.51 \\                                         
\hline                                                                                       
\multirow{4}*{$\sigma\pi-2elec$} & c0 & 14.20 & 0.36 & 11.46 & 0.38 & 45.68 & 0.55 \\        
 & c1 & 10.67 & 0.32 & 14.68 & 0.36 & 62.98 & 0.66 \\                                        
 & c3 & 19.19 & 0.34 & 8.60 & 0.28 & 79.26 & 0.75 \\                                         
 & c3 & 17.85 & 0.30 & 8.11 & 0.27 & 74.10 & 0.71 \\                                         
\hline                                                                                       
\multirow{4}*{JK} & c0 & 4.61 & 0.21 & 5.60 & 0.23 & 8.89 & 0.47 \\                          
 & c1 & 4.43 & 0.21 & 5.47 & 0.23 & 8.46 & 0.41 \\                                           
 & c3 & 4.31 & 0.21 & 5.18 & 0.22 & 7.94 & 0.46 \\                                           
 & c3 & 4.15 & 0.20 & 5.21 & 0.22 & 7.81 & 0.42 \\                                           
\hline                                                                                       
\multirow{4}*{$\sigma\pi$-JK} & c0 & 5.58 & 0.23 & 6.10 & 0.25 & 9.68 & 0.37 \\              
 & c1 & 4.50 & 0.21 & 5.29 & 0.22 & 7.88 & 0.23 \\                                           
 & c3 & 7.04 & 0.23 & 5.29 & 0.22 & 7.88 & 0.23 \\                                           
 & c3 & 4.25 & 0.20 & 5.17 & 0.22 & 7.68 & 0.24 \\                                           
\hline                                                                                       
\end{tabular} 
\caption{RMS errors for models trained on the \textit{ethylene} dataset with inclusion of decomposed energies. Units are $\SI{}{\electronvolt}$ for $E_{orb}$ and $\SI{}{\kcal\per\mole}$ for $E_{tot}$.}                   \label{tab:Ethylprop2trans}       
\end{table}

\begin{table}[h]                                                                            
\centering                                                                                   
\begin{tabular}{cccccccc}                                                                    
\hline                                                                                       
 &  & \multicolumn{2}{c}{Train} & \multicolumn{2}{c}{Validation} & \multicolumn{2}{c}{Propylene} \\
 &  & E$_{tot}$ & E$_{orb}$ & E$_{tot}$ & E$_{orb}$ & E$_{tot}$ & E$_{orb}$ \\               
\hline                                                                                       
 & Initial & 20.95 & 2.88 & 18.98 & 2.87 & 20.03 & 2.78 \\                                   
\hline                                                                                       
\multirow{4}*{2elec} & c0 & 9.62 & 0.28 & 10.78 & 0.34 & 79.86 & 0.58 \\                     
 & c1 & 4.23 & 0.18 & 4.63 & 0.19 & 93.62 & 0.52 \\                                          
 & c2 & 4.14 & 0.18 & 4.62 & 0.19 & 97.27 & 0.52 \\                                          
 & c3 & 4.58 & 0.19 & 4.62 & 0.19 & 97.27 & 0.52 \\                                          
\hline                                                                                       
\multirow{4}*{$\sigma\pi-2elec$} & c0 & 4.45 & 0.20 & 4.96 & 0.22 & 5.88 & 0.39 \\           
 & c1 & 4.13 & 0.19 & 5.23 & 0.21 & 5.94 & 0.38 \\                                           
 & c3 & 3.90 & 0.18 & 5.25 & 0.20 & 5.33 & 0.35 \\                                           
 & c3 & 3.88 & 0.18 & 5.28 & 0.20 & 5.38 & 0.37 \\                                           
\hline                                                                                       
\multirow{4}*{JK} & c0 & 5.27 & 0.20 & 6.29 & 0.22 & 7.28 & 0.34 \\                          
 & c1 & 3.71 & 0.18 & 5.48 & 0.22 & 5.81 & 0.35 \\                                           
 & c3 & 3.59 & 0.19 & 5.06 & 0.21 & 5.63 & 0.35 \\                                           
 & c3 & 3.58 & 0.18 & 5.06 & 0.21 & 5.63 & 0.35 \\                                           
\hline                                                                                       
\multirow{4}*{$\sigma\pi$-JK} & c0 & 3.72 & 0.16 & 5.25 & 0.20 & 6.42 & 0.30 \\              
 & c1 & 3.48 & 0.16 & 4.97 & 0.20 & 6.37 & 0.32 \\                                           
 & c3 & 3.57 & 0.15 & 4.93 & 0.19 & 6.00 & 0.28 \\                                           
 & c3 & 3.41 & 0.16 & 4.90 & 0.19 & 5.97 & 0.28 \\                                           
\hline                                                                                       
\end{tabular}                                               
\caption{RMS errors for models trained on the \textit{ethylene} dataset without inclusion of decomposed energies. Units are $\SI{}{\electronvolt}$ for $E_{orb}$ and $\SI{}{\kcal\per\mole}$ for $E_{tot}$.}
\label{tab:Ethylprop1trans} 
\end{table}

\begin{table}[h]                                                   
\centering                                                         
\begin{tabular}{cccccccccccccc}                                                   
\hline                                     
 &  & \multicolumn{2}{c}{KE} & \multicolumn{2}{c}{EN$_H$} & \multicolumn{2}{c}{EN$_A$} & \multicolumn{2}{c}{E2} & \multicolumn{2}{c}{E$_{tot}$} & \multicolumn{2}{c}{E$_{orb}$} \\
 & Context & train & validation & train & validation & train & validation & train & validation & train & validation & train & validation \\ 
\hline
Initial & & 167.04 & 157.99 & 27.96 & 26.66 & 405.17 & 389.89 & 140.60 & 136.62 & 21.47 & 19.21 & 4.44 & 4.43 \\
\hline                                                                  
\multirow{4}*{2elec} & c0 & 97.67 & 100.54 & 26.33 & 25.00 & 233.60 & 223.97 & 90.79 & 93.07 & 6.83 & 6.65 & 0.29 & 0.30 \\ 
 & c1 & 40.04 & 42.03 & 14.28 & 13.36 & 153.40 & 145.59 & 65.12 & 66.21 & 3.78 & 4.05 & 0.18 & 0.18 \\ 
 & c2 & 39.91 & 41.74 & 14.31 & 13.38 & 152.70 & 145.12 & 64.75 & 65.64 & 3.98 & 4.03 & 0.18 & 0.18 \\ 
 & c3 & 39.90 & 41.89 & 13.05 & 12.01 & 154.75 & 137.57 & 60.47 & 63.62 & 3.39 & 3.83 & 0.18 & 0.18 \\ 
\hline 
\multirow{4}*{$\sigma\pi-2elec$} & c0 & 92.95 & 93.10 & 28.09 & 26.58 & 243.83 & 232.82 & 93.16 & 96.67 & 6.32 & 6.49 & 0.29 & 0.29 \\ 
 & c1 & 42.60 & 41.52 & 18.53 & 17.32 & 178.48 & 161.34 & 65.15 & 62.55 & 3.78 & 4.01 & 0.18 & 0.18 \\ 
 & c2 & 40.29 & 38.58 & 16.77 & 15.52 & 155.84 & 143.03 & 57.86 & 57.09 & 3.36 & 3.82 & 0.17 & 0.19 \\ 
 & c3 & 39.25 & 37.37 & 16.32 & 15.04 & 159.95 & 141.90 & 57.05 & 56.91 & 3.08 & 3.83 & 0.17 & 0.18 \\ 
\hline 
\multirow{4}*{JK} & c0 & 94.00 & 96.75 & 29.51 & 28.20 & 251.62 & 252.49 & 102.11 & 107.49 & 7.39 & 7.64 & 0.37 & 0.33 \\        
 & c1 & 90.92 & 94.36 & 29.54 & 28.25 & 260.68 & 261.02 & 100.31 & 106.59 & 5.05 & 5.45 & 0.31 & 0.31 \\
 & c2 & 46.08 & 49.13 & 16.60 & 15.80 & 171.46 & 162.33 & 63.43 & 68.91 & 3.78 & 4.37 & 0.25 & 0.27 \\  
 & c3 & 44.43 & 47.38 & 16.01 & 15.32 & 157.85 & 161.04 & 63.33 & 68.63 & 3.70 & 4.05 & 0.24 & 0.25 \\  
\hline  
\multirow{4}*{$\sigma\pi$-JK} & c0 & 91.61 & 93.91 & 27.51 & 26.18 & 250.70 & 240.54 & 90.61 & 94.27 & 7.99 & 8.45 & 0.27 & 0.28 \\        
 & c1 & 58.18 & 60.26 & 24.26 & 23.05 & 201.23 & 189.47 & 67.30 & 69.29 & 3.90 & 4.56 & 0.24 & 0.25 \\ 
 & c2 & 53.23 & 55.08 & 23.07 & 21.93 & 191.30 & 180.63 & 65.12 & 67.22 & 3.52 & 4.26 & 0.23 & 0.24 \\ 
 & c3 & 53.12 & 55.01 & 23.07 & 21.93 & 188.73 & 181.06 & 65.13 & 67.43 & 3.51 & 4.23 & 0.22 & 0.24 \\ 
\hline                                        
\end{tabular}
\caption{RMS errors for models trained on the \textit{combined} dataset with inclusion of decomposed energies. Units are $\SI{}{\electronvolt}$ for $E_{orb}$ and $\SI{}{\kcal\per\mole}$ for all other quantities.}
\label{tab:EEprop2} 
\end{table}

\begin{table}[h]                                 
\centering                
\begin{tabular}{cccccccccccc}             
\hline
&  & \multicolumn{2}{c}{Train} & \multicolumn{2}{c}{Validation} & \multicolumn{2}{c}{Propane} & \multicolumn{2}{c}{t-Butane} & \multicolumn{2}{c}{Propylene} \\
 & Context & E$_{tot}$ & E$_{orb}$ & E$_{tot}$ & E$_{orb}$ & E$_{tot}$ & E$_{orb}$ & E$_{tot}$ & E$_{orb}$ & E$_{tot}$ & E$_{orb}$ \\   
\hline                                            
 Initial & & 21.47 & 4.44 & 19.21 & 4.43 & 26.68 & 5.38 & 32.69 & 5.31 & 20.03 & 2.78 \\
\hline                              
\multirow{4}*{\textit{2elec}} & c0 & 6.83 & 0.29 & 6.65 & 0.30 & 9.46 & 0.57 & 10.36 & 0.55 & 5.08 & 0.24 \\
 & c1 & 3.78 & 0.18 & 4.05 & 0.18 & 4.42 & 0.32 & 7.50 & 0.31 & 524.89 & 1.78 \\
 & c2 & 3.98 & 0.18 & 4.03 & 0.18 & 4.38 & 0.32 & 7.44 & 0.31 & 580.27 & 1.94 \\
 & c3 & 3.39 & 0.18 & 3.83 & 0.18 & 3.95 & 0.30 & 751.20 & 1.82 & 627.40 & 2.03 \\
\hline  
\multirow{4}*{$\sigma\pi$-2elec} & c0 & 6.32 & 0.29 & 6.49 & 0.29 & 8.18 & 0.53 & 10.93 & 0.49 & 6.69 & 0.27 \\                                           
 & c1 & 3.78 & 0.18 & 4.01 & 0.18 & 5.14 & 0.31 & 1105.67 & 1.41 & 851.01 & 1.60 \\ 
 & c3 & 3.36 & 0.17 & 3.82 & 0.19 & 4.82 & 0.29 & 4234.52 & 4.28 & 1743.58 & 4.06 \\
 & c3 & 3.08 & 0.17 & 3.83 & 0.18 & 4.82 & 0.31 & 4531.66 & 4.79 & 2109.05 & 4.36 \\
\hline                                                                              
\multirow{4}*{\textit{JK}} & c0 & 7.39 & 0.37 & 7.64 & 0.33 & 10.38 & 0.39 & 15.66 & 0.50 & 8.30 & 0.27 \\
 & c1 & 5.05 & 0.31 & 5.45 & 0.31 & 9.18 & 0.38 & 14.39 & 0.47 & 6.71 & 0.22 \\
 & c3 & 3.78 & 0.25 & 4.37 & 0.27 & 5.71 & 0.32 & 9.18 & 0.44 & 5.60 & 0.28 \\ 
 & c3 & 3.70 & 0.24 & 4.05 & 0.25 & 5.67 & 0.31 & 9.19 & 0.42 & 5.72 & 0.28 \\ 
\hline 
\multirow{4}*{$\sigma\pi$-\textit{JK}} & c0 & 7.99 & 0.27 & 8.45 & 0.28 & 12.92 & 0.36 & 17.86 & 0.48 & 11.85 & 0.22 \\                                            
 & c1 & 3.90 & 0.24 & 4.56 & 0.25 & 6.42 & 0.37 & 11.84 & 0.42 & 7.39 & 0.20 \\
 & c3 & 3.52 & 0.23 & 4.26 & 0.24 & 5.54 & 0.36 & 11.39 & 0.40 & 6.28 & 0.22 \\
 & c3 & 3.51 & 0.22 & 4.23 & 0.24 & 5.49 & 0.36 & 11.40 & 0.40 & 6.25 & 0.21 \\
\hline                                   
\end{tabular}             
\caption{RMS errors for models trained on the \textit{combined} dataset with inclusion of decomposed energies. Units are $\SI{}{\electronvolt}$ for $E_{orb}$ and $\SI{}{\kcal\per\mole}$ for all other quantities.}
\label{tab:EEprop2transSP}
\end{table} 

\begin{table}[h]                                      
\centering                                               
\begin{tabular}{cccccccccccc}                                  
\hline                                          
 &  &\multicolumn{2}{c}{Train} &\multicolumn{2}{c}{Validation} & \multicolumn{2}{c}{Propane} & \multicolumn{2}{c}{t-Butane} & \multicolumn{2}{c}{Propylene}  \\
 & Context & E$_{tot}$ & E$_{orb}$ & E$_{tot}$ & E$_{orb}$ & E$_{tot}$ & E$_{orb}$ & E$_{tot}$ & E$_{orb}$ & E$_{tot}$ & E$_{orb}$ \\
\hline 
Initial & & 21.47 & 4.44 & 19.21 & 4.43 & 26.68 & 5.38 & 32.69 & 5.31 & 20.03 & 2.78 \\
\hline                                          
\multirow{4}*{2elec} & c0 & 5.86 & 0.25 & 6.07 & 0.26 & 6.64 & 0.39 & 8.58 & 0.44 & 4.04 & 0.26 \\              
 & c1 & 3.05 & 0.15 & 3.63 & 0.17 & 3.96 & 0.34 & 7.44 & 0.63 & 4.99 & 0.17 \\
 & c2 & 3.08 & 0.15 & 3.62 & 0.17 & 3.97 & 0.34 & 7.48 & 0.65 & 4.05 & 0.26 \\
 & c3 & 3.00 & 0.14 & 3.52 & 0.17 & 3.84 & 0.35 & 7.18 & 0.66 & 3.90 & 0.26 \\
\hline                                                                            
\multirow{4}*{$\sigma\pi$-2elec} & c0 & 5.33 & 0.26 & 15.09 & 0.31 & 7.10 & 0.38 & 11.50 & 0.45 & 5.93 & 0.27 \\ 
 & c1 & 3.28 & 0.15 & 3.94 & 0.18 & 4.60 & 0.31 & 10.32 & 0.55 & 311.65 & 1.17 \\
 & c2 & 3.27 & 0.14 & 3.66 & 0.17 & 4.65 & 0.44 & 1514.45 & 2.28 & 740.76 & 1.74 \\
 & c3 & 2.66 & 0.13 & 3.59 & 0.16 & 4.44 & 0.50 & 2046.02 & 2.24 & 255.79 & 1.20 \\ 
\hline  
\multirow{4}*{\textit{JK}}  & c0 & 4.49 & 0.23 & 5.35 & 0.24 & 6.31 & 0.41 & 9.34 & 0.58 & 5.23 & 0.36 \\                 
 & c1 & 3.68 & 0.20 & 4.45 & 0.21 & 5.47 & 0.33 & 8.49 & 0.48 & 4.91 & 0.27 \\ 
 & c2 & 3.56 & 0.20 & 4.28 & 0.20 & 5.19 & 0.32 & 8.21 & 0.48 & 4.76 & 0.26 \\ 
 & c3 & 3.70 & 0.20 & 4.29 & 0.20 & 5.12 & 0.33 & 8.13 & 0.48 & 4.72 & 0.26 \\ 
\hline  
\multirow{4}*{$\sigma\pi$\textit{-JK}} & c0 & 5.77 & 0.24 & 6.33 & 0.26 & 10.64 & 0.40 & 15.75 & 0.44 & 9.13 & 0.20 \\     
 & c1 & 4.03 & 0.20 & 4.68 & 0.21 & 7.55 & 0.36 & 13.89 & 0.40 & 6.93 & 0.19 \\ 
 & c2 & 4.09 & 0.20 & 4.72 & 0.21 & 7.47 & 0.36 & 13.82 & 0.40 & 6.85 & 0.19 \\ 
 & c3 & 4.20 & 0.21 & 4.62 & 0.21 & 7.36 & 0.36 & 13.70 & 0.40 & 6.79 & 0.19 \\ 
\hline                              
\end{tabular}
\caption{RMS errors for models trained on the \textit{combined} dataset without inclusion of decomposed energies. Units are $\SI{}{\electronvolt}$ for $E_{orb}$ and $\SI{}{\kcal\per\mole}$ for all other quantities.} 
\label{tab:EEprop1transSP}  
\end{table}

\end{document}